\newif\iffigures
 	\figurestrue  

\newif\ifMPIfR  
 \MPIfRfalse  

\ifMPIfR
	\iffigures
		\documentstyle[times,rotate,psfig]{l-aa}
	\else
		\documentstyle[times,rotate,psfig]{l-aa}
	\fi
\else
	\iffigures
		\documentstyle[rotate,psfig]{l-aa}
	\else
		\documentstyle[rotate]{l-aa} 
	\fi
\fi			

\newlength{\colwidth}
\begin{document}

\thesaurus{02.18.5; 03.09.1; 11.02.1;
12.04.2; 13.07.3}

\title{Beacons at the Gamma Ray Horizon}

\author{K.~Mannheim\inst{1}
\and S.~Westerhoff\inst{2}
\and H.~Meyer\inst{2}
\and H.-H.~Fink\inst{3}}

\institute{Universit\"ats-Sternwarte, Geismarlandstr.\ 11, D-37083
G\"ottingen, Germany
\and Universit\"at Wuppertal, Fachbereich Physik, D-42097 Wuppertal, Germany
\and Max-Planck-Institut f\"ur extraterrestrische Physik, D-85740 Garching, 
Germany} 
\offprints{K.~Mannheim (kmannhe@uni-sw.gwdg.de)}
\date{received date; accepted date}
\maketitle

\begin{abstract}

Blazars with redshifts $z\le 0.1$ are likely candidates for
detection at energies in the range 300~GeV -- 50~TeV with $\rm \check C$erenkov telescopes
and scintillator arrays.   We present $\gamma$-ray flux predictions  
for a sample of 15 nearby flat-spectrum radio sources  
fitting the proton blazar model of Mannheim (1993a) 
to their observed broad-band spectral
energy distributions.    At high energies, we use fluxes or flux limits
measured by
ROSAT,  CGRO and the Whipple Observatory to constrain their spectra. 
We take into account absorption of the $\gamma$-rays by 
pair production with low energy photons of the diffuse infrared-to-optical photon background
produced by galaxies (cosmic absorption) and with low energy synchrotron photons of
the blazar radiation field
(internal absorption).  
Typically, the theoretical spectra decrease much faster above TeV 
(photon index $s\approx  3$) than between GeV and TeV ($s\approx 2$) 
owing to internal absorption.\\
The predicted fluxes are confronted with flux limits in the 20-50~TeV energy range
obtained by the {\it High Energy Gamma Ray Astronomy} (HEGRA) experiment.
Without cosmic absorption, the fluxes are 
about equal to the current sensitivity of HEGRA.
Improved $\gamma$/hadron separation techniques could render 
a detection by HEGRA possible, if cosmic absorption by the far-infrared
background at wavelengths $\sim\!100$~$\mu$m
is not exceedingly strong.  
 
\end{abstract}

\section{Introduction}
\label{intro}
One of the most remarkable results obtained with
the spark chamber experiment EGRET onboard
the Compton Gamma Ray Observatory is the discovery
of a large number of $\gamma$-ray point sources at high
galactic latitudes (Fichtel et al. \cite{fichtel94}, Montigny et al. 
\cite{montigny95}).  
Most of the sources have
been identified as blazars, which comprise an important
subclass of active galactic nuclei (AGN).  Blazars 
are characterized by highly variable,
featureless and polarized continuum emission.  In the 
radio-to-soft X-ray range, the emission is
commonly attributed to synchrotron emission from a
collimated plasma jet (Blandford \& K\"onigl
\cite{blandford79}) presumably emerging
from the rotating magnetosphere of an accreting supermassive
black hole of mass $M\sim 10^8$-$10^{10}M_\odot$ (Camenzind \cite{camenzind90}).
Due to relativistic bulk motion,
the radiation pattern of the jet
is sharply peaked about its axis.
In blazars, the jet axis is aligned closely with the line-of-sight,
leading to Doppler-boosted emission and superluminal motion
of radio knots.\\
Recent multifrequency observations of blazars
(Maraschi et al. \cite{maraschi94}, 
Falomo et al., \cite{falomo95},
Valtaoja \& Ter\"asranta \cite{valtaoja95}) show that mm, 
optical, X-ray and $\gamma$-ray lightcurves all
trace the same event during a `flare'.  This
favors a synchrotron-self-Compton (SSC) 
or proton-initiated-cascade
(PIC, Mannheim et al. \cite{mannheim91}) origin of the $\gamma$-rays at evolving shock fronts.
Both mechanisms predict correlated TeV emission (Zdziarski \& Krolik
\cite{zdziarski94}, Mannheim \cite{mannheim96}). \\ 
\begin{figure} 
\newcommand{\picturebox}{\framebox[8.7cm]{\rule{0pt}{6.9cm}}}
\iffigures
\centerline{\psfig{figure=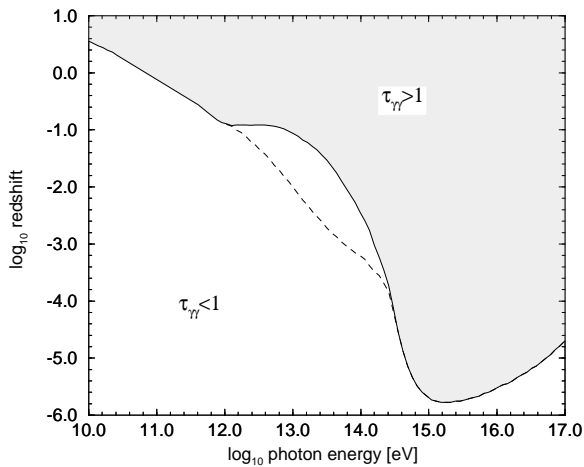,height=7cm}} 
\else     
\centerline{\picturebox}
\fi
\caption[]{\label{pic_horizon} 
The $\gamma$-ray horizons $\tau_{\gamma\gamma}(E,z)=1$
corresponding to the two different diffuse background models shown in Fig.2.
The horizons were calculated assuming $\Omega=1$, $q_\circ=0.5$,
$H_\circ=75$~km~s$^{-1}$~Mpc$^{-1}$ and a photon number
density evolving with redshift as $n'd\epsilon'=(1+z)^3nd\epsilon$
(conserved number of photons)}
\end{figure}
At high photon energies, the $\gamma$-radiation from remote cosmic sources is severely
attenuated by pair production via $\gamma+\gamma\rightarrow e^++e^-$ on low energy
background photons
(Gould \& Shr\`eder \cite{gould66}).
Stecker et al. (\cite{stecker92}) propose
to infer the diffuse near-infrared background radiation,
which is difficult to observe directly, 
by measuring the expected TeV cutoff in the spectra of blazars at $z\approx 0.1$
owing to cosmic absorption.
Salamon et al. (\cite{salamon94}) argue that 
$\gamma$-ray measurements could also be used as a yardstick to
determine the cosmic distance scale.
MacMinn \& Primack (\cite{macminn96}) point out that the diffuse
infrared-to-ultraviolet background is directly related to
models of galaxy formation.  Measuring the diffuse background
from the infrared to the ultraviolet bands, requires  
observations of blazars at $\gamma$-ray energies from 10~GeV
(Madau \& Phinney \cite{madau96}) to 50~TeV.
Probing the Universe with $\gamma$-rays will allow for independent
tests of cosmological scenarios based on the well-understood 
physics of pair production.\\
Figure 1 shows the $\gamma$-ray horizon marking the  
relation between  
photon energy $E$ 
and redshift $z$ of a $\gamma$-ray source  
defined by $\tau_{\gamma\gamma}(E,z)=1$.
The solid curve is obtained by numerically integrating the  
optical depth $\tau_{\gamma\gamma}$ as given in 
Dwek \& Slavin (\cite{dwek94}) using the  present-day
background spectrum computed by averaging various galactic evolutionary scenarios
of MacMinn \& Primack (\cite{macminn96}) shown in Fig.2.  
This spectrum remains well below current experimental
limits for the near-infrared background  (Biller et al. \cite{biller95}).
Practically, knowledge of the
turnover energy $E$ for a given redshift would yield the  
value of the diffuse background photon density above the
resonant energy $\epsilon=2m_{\rm e}c^2/[E(1+z)^2]$.  
Compared to our standard scenario 
for $E(z)$ shown as the solid line in Fig.~1,
the turnover energy increases if (i)  the Hubble constant 
has a value greater than $75$~km~s$^{-1}$~Mpc$^{-1}$, (ii) the  
diffuse background radiation has a density lower than the value obtained
by averaging the MacMinn \& Primack models or (iii)
the evolution of the diffuse background density $n'(\epsilon')d\epsilon'\propto
(1+z)^{\rm k}$ is more shallow than $k=3$ already for moderate redshifts
(keeping $\Omega=1$ and  $\Lambda=0$ fixed). 
A low background photon density would indicate late galaxy formation,
and thus only a small amount of cold dark matter in the models
of MacMinn \& Primack. 
Effects (i)--(iii) can reduce the optical depth $\tau_{\gamma\gamma}(E,z)$
by a maximum factor of $\sim 3$ below the values adopted in the
present work assuming the extremal values
$H_\circ=100$~km~s$^{-1}$~Mpc$^{-1}$, a present-day
infrared background equal to the integrated number counts of IRAS
galaxies (de Zotti et al. \cite{dezotti95}) and $k=0$.  \\
\begin{figure} 
\newcommand{\picturebox}{\framebox[8.7cm]{\rule{0pt}{6.9cm}}}
\iffigures
\centerline{ \psfig{figure=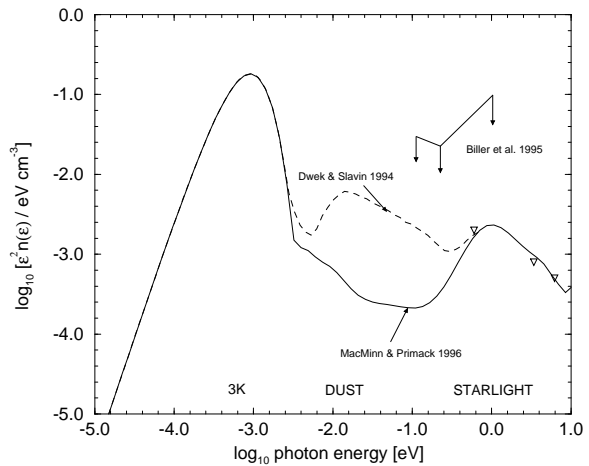,height=7cm}} 
\else     
\centerline{\picturebox}
\fi
\caption[]{\label{pic_cibr}  {\it Solid line:}
the infrared-to-ultraviolet diffuse
background radiation field adopted in the present work.
{\it Dashed line:} a diffuse background assuming that the $\gamma$-ray spectrum
of Mrk421 cuts off at TeV due to cosmic absorption.  Triangles denote estimates
by Madau \& Phinney (\cite{madau96}) 
of the optial-to-ultraviolet diffuse background based on deep galaxy
surveys }
\end{figure}By inspection of Fig.~1 it is obvious that sources at redshifts 
less than $\sim 0.1$  
are required to pin down
the $\gamma$-ray horizon in the TeV range.  
Several low-redshift EGRET blazars have been found, 
of which Mrk421 and Mrk501 have also been detected
at TeV energies (Punch et al. \cite{punch92}, Quinn et al. \cite{quinn96}),
and many more potential sources with similar overall
spectrum exist. To extract
information on the intergalactic absorption from 
observed TeV spectra requires detailed knowledge of the
spectra intrinsic to the sources.\\
In this paper, we therefore investigate  
a sample of nearby flat-spectrum radio sources 
by fitting the proton blazar model of Mannheim (\cite{mannheim93})
to their multifrequency spectra. 
In Sect.2  we briefly outline the model and explain the 
fitting procedure.
In Sect.3, we introduce the sample
and proceed in Sect.4 comparing the predicted fluxes above TeV
with the sensitivity of  
HEGRA.
Finally, we summarize our results on the feasibility
of mapping the $\gamma$-ray horizon in the TeV range.

\section{Outline of the proton-initiated cascade model}
Recent multifrequency observations of blazars seem to support an explanation of
their $\gamma$-ray emission in terms of relativistic shocks propagating
down an expanding jet (at least this is a powerful 
and predictive working hypothesis).    Multiple shocks (Courant \& Friedrichs
\cite{courant48})
can form
behind a magnetic nozzle (van Putten \cite{putten95}). 
Particle acceleration at relativistic and oblique shocks 
by diffusive and drift mechanisms has been studied by
Kirk \& Heavens (\cite{kirk89}), Schneider \& Kirk (\cite{schneider89}),
Kirk \& Schneider (\cite{kirks89}), Webb (\cite{webb89}), 
Begelman \& Kirk (\cite{begelman90}), Kr\"ulls (\cite{kruells90}),
Kirk (\cite{kirk92}),  Ellison et al. (\cite{ellison95}).
In the innermost regions of the jet,
at the beginning of an outburst, the shocks seem to emit an optically
thin mm-to-optical synchrotron spectrum  (in some cases reaching up
to soft X-rays) with an associated X-to-$\gamma$-ray spectral component
which can dominate the emitted power.
Further down the jet, the shocks become transparent to radio synchrotron
emission so that the initial outburst shows up delayed and broadened in radio lightcurves.
Morphologically, the radio outbursts seem to be associated with
the birth of new knots in VLBI maps.\\
It is tantalizing to identify the high energy spectral component
with the inevitable
synchrotron-self-Compton emission from the traveling shocks.
However, this would require that the photon energy density  exceeds
the magnetic energy density in sources where the $\gamma$-ray luminosity exceeds
the optical luminosity, i.e. $u_{\rm rad}\gg u_{\rm B}$, since
\begin{equation}
L_{\rm ssc}\simeq {u_{\rm rad}\over   u_{\rm B}} L_{\rm syn}
\end{equation}
The same conclusion  
$u_{\rm rad}\gg u_{\rm B}$  
follows from assuming that TeV photons are singly scattered
optical synchrotron photons.   Since $x_{\rm opt}=h\nu/(m_{\rm e}c^2)=
3.4\, 10^{-14}B_\perp \gamma^2$ and $x_{\rm tev}={4\over 3}x_{\rm opt}\gamma^2$,
we obtain $B_\perp=3\, 10^{13}x_{\rm opt}^2/x_{\rm tev}~{\rm G}\approx 3\, 10^{-3}$~G
corresponding to the comoving-frame magnetic energy density of 
$\sim 4\, 10^{-7}\gamma_{\rm j}^{-2}$erg~cm$^{-3}$.  This value is negligible compared to the
comoving-frame radiation energy density $\sim 1$~erg~s$^{-1}$
(Sect.3.2).
However, the energy density of the relativistic electrons $u_{\rm rel}$
should certainly exceed the energy density $u_{\rm rad}$ of the photons they produce
implying $u_{\rm rel}\gg u_{\rm B}$.
This raises the problem of how $u_{\rm rel}$ can be maintained at such a high level
in a quasi-stationary situation.  The commonly invoked mechanism of diffusive
acceleration does not operate at $u_{\rm rel}\gg u_{\rm B}$, since
particles diffuse by pitch-angle scattering off magnetic field
fluctuations.   The pressure of these fluctuations is expected to
be near the saturation value $u_{\delta\rm B}\approx u_{\rm B}$, thus
imposing the condition $u_{\rm rel}\le u_{\rm B}$.\\
This conceptual problem with the SSC mechanism is removed if
it is assumed that protons instead of electrons
scatter off the synchrotron photons.  In interstellar space,
the energy in relativistic protons largely exceeds the energy in electrons
which could also be true in radio jets.   The reasons for this are unclear.
However, if the proton acceleration mechanism 
can tap the Maxwell tail of a thermal distribution (Malkov \& V\"olk \cite{malkov95}),
it is plausible that the proton energy density will generally be larger than
the electron energy density.
Protons in a high
radiation density environment, such as the  
shocks presumably traveling down radio jets, produce ample
pions and pairs by inelastic scattering off the electronic synchrotron
photons if they reach high energies  $\sim
10^{8-11}$~GeV (Biermann \& Strittmatter 1987).\\
The pairs and pions initiate electromagnetic cascades 
(Mannheim et al. 1991) at the shocks.  For a simple
Blandford \& K\"onigl (1979) type jet, the predicted
spectra match the observed
spectra from radio frequencies to $\gamma$-rays very well
(Mannheim 1993a).   
Contrasting any other model predicting
$\gamma$-rays from blazars, the proton blazar model implies
source spectra reaching $>$~TeV photon energies {\it generally}.
Internal absorption of the $\gamma$-rays above TeV by the low energy electronic
synchrotron photons
leads to a steepening of the spectra by one power 
above TeV ($s\approx 3$).  
The PIC flux varies simultaneously with the flux of the infrared-to-optical
target
photons and with the proton maximum energy. \\
\begin{figure} 
\newcommand{\picturebox}{\framebox[8.7cm]{\rule{0pt}{6.5cm}}}
\iffigures
\centerline{\psfig{figure=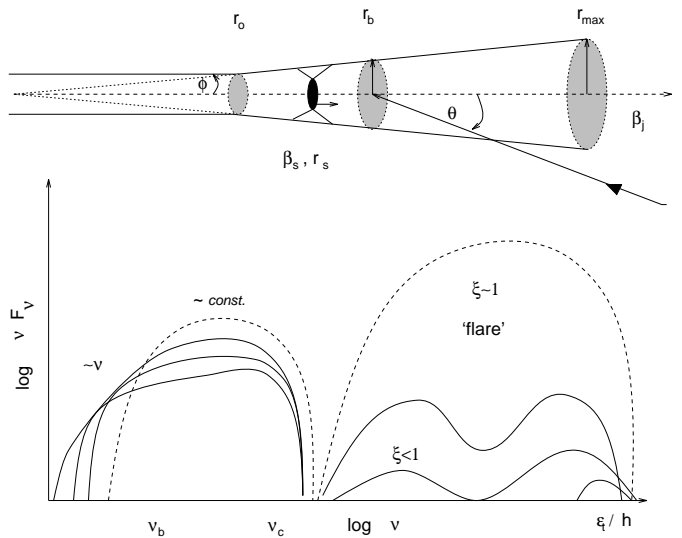,height=7cm}} 
\else     
\centerline{\picturebox}
\fi
\caption[]{\label{pic_sketch} 
Sketch of  jet geometry and typical synchrotron + PIC spectrum.
{\it Solid lines} show the emission from various $r$ in the
range $r_{\rm b}\le r\le r_{\rm max}$.  {\it Local} 
electronic synchrotron spectra
are characterized with increasing frequency
by a synchrotron-self-absorption turnover, an optical thin
$S_\nu\propto \nu^{-1/2}$ radio/mm spectrum, a $S_\nu\propto \nu^{-1}$ 
infrared spectrum and a steep near-infrared/optical/UV spectrum.
The superposition of many such spectra along the expanding jet yields 
a flat radio spectrum with $S_\nu \propto const.$   
and a $S_\nu\propto \nu^{-1}$ spectrum above the break frequency
$\nu_{\rm b}$ turning over steeply
at $\nu_{\rm c}$.   The corresponding PIC spectrum rises roughly as
$S_\nu\propto \nu^{-(0.5-0.9)}$  (depending on the pair
creation optical depth $\tau_{\gamma\gamma}$) and steepens in the
MeV to GeV range where $S_\nu\propto \nu^{-1}$ is typical.
Above $\sim \rm TeV$, the PIC spectra steepen further by one power and then 
cut-off at ultra high energies.  {\it Dashed lines} indicate
the spectrum emitted when a newly formed shock ($\xi\sim 1$)
enters the region
at $r= r_{\rm b}$ where the target density for proton cooling rises sharply
(the synchrotron radiation at target frequencies becomes
optically thin at this point).  Further down the jet, the $\gamma$-ray
luminosity decreases, since the proton maximum energy required
for $\xi\sim 1$ cannot be maintained by the 
nonlinearly evolving shocks
}
\end{figure}
Denoting the emission volume as $V$ (assumed to be equal
for protons and electrons), the energy density ratio of 
protons and electrons as $\eta=u_{\rm p}/u_{\rm e}$  
($\eta\approx 100$ in the Milky Way at GeV energies) and 
the ratio of electron and proton cooling times 
as $\xi=t_{\rm e}/t_{\rm p}$, 
the PIC luminosity can be expressed in terms of
the SSC luminosity Eq.(1) by
\begin{equation}
L_{\rm pic}\simeq u_{\rm p}V t_{\rm p}^{-1}=\eta u_{\rm e}Vt_{\rm p}^{-1} 
\simeq \eta L_{\rm syn}t_{\rm e}t_{\rm p}^{-1}=\eta\xi L_{\rm syn}
\end{equation}
The cooling time ratio $\xi$
depends strongly on the maximum particle energies, since $t_{\rm cool}\propto
1/\gamma$ where $\gamma$ denotes the Lorentz factor of the particle.
Balancing cooling and expansion time scales for
protons and electrons (so that $\xi=1$) yields 
extremely large proton Lorentz factors $\gamma_{\rm p}\sim 10^{8-11}$.
In the case of $\xi=1$,
the PIC luminosity can exceed the SSC luminosity by the, possibly large, 
proton-to-electron ratio $\eta$.
A crucial requirement is, of course, that acceleration is faster than expansion
and that the Larmor radius of the gyrating particles does not
exceed the radius of curvature of the shocks (Bell 1978).
The latter criterion hurts protons more than electrons, since
the proton Larmor radius is much larger than the electron Larmor radius.
If the coherent shock structure is destroyed during the nonlinear
evolution of the shock propagating down the jet, proton scattering at the maximum
energy becomes inefficient (scattering length $\sim$ Larmor radius
in Bohm diffusion). 
We therefore expect that
$\xi(t)\propto\gamma_{\rm p,max}/\gamma_{\rm e,max}$ is
decreasing with time.
Multiple  shocks  generate some time-average $\langle\xi\rangle$ which will be relevant
for our sample of randomly observed
blazars.\\ 
\begin{table*}
\caption{
Summary of high-energy data available for northern blazars
with $z\le 0.1$ or unkown redshifts.  The numbers in the columns below the
experiments give the observed differential energy flux 
or flux limit $\log_{10}[\rm E^2dN/dE\ (erg~cm^{-2}~s^{-1})]$ at the 
threshold energy of the respective experiment}
\begin{minipage}{\textwidth}
\renewcommand{\footnoterule}{\rule{0pt}{0pt}}
\renewcommand{\thefootnote}{\arabic{mpfootnote}}
\label{table_sources}
\begin{tabular}{llccrrcc}
&\bf Source &\bf Redshift &\bf
ROSAT\footnote{PSPC differential flux at 1~keV
adopting $\alpha=0.5$ ($S_\nu\propto \nu^{-
\alpha}$) }
& \bf CGRO\footnote{EGRET differential flux at 100~MeV adopting $\alpha=1$;
Fichtel et al. (1994), Thompson et al. 1995, Montigny et al.
(1995)\\ \indent \ \
For BL~Lac, we give a preliminary EGRET flux  
(Hartman 1995)} 
& \bf Whipple\footnote{$\rm \check  C$erenkov telescope data
at a threshold energy of $0.35$~TeV adopting 
$\alpha=2$;  Kerrick et al. (\cite{kerrick95a})} 
&\bf  HEGRA\footnote{
Scintillator array data at the threshold $40$~TeV
adopting an exponentially decreasing  spectrum \\ \indent \ \ 
For Mrk~421 and Mrk~501 we also give the AIROBICC data for $E_{\rm th}\simeq 25$~TeV 
\\ \indent \ \ 
Data are from K\"uhn (\cite{kuehn94}), Karle et al. (\cite{karle95}) 
and Westerhoff
et al. (1995b) on behalf of the HEGRA collaboration}
&\bf Tibet\footnote{Differential flux from air shower data at $10$~TeV
adopting an exponential spectrum; Amenomori et al. (1994)}\\
%
\hline
&  0116+319 (MS 01166+31)    &    0.059   & $-12.3 $    &  $<-10.9$    & $<-10.9$ & $<-11.0$   & -- \\ 
&  0430+052  (3C120.0)       &    0.033   & $-10.2$     & $<-10.7$     & --           & $<-10.5 $ & -- \\ 
&  0716+714                  &     --     & $-11.6$     & $-10.5$      & $<-10.8$  & --            & --  \\ 
&  0802+243  (3C192.0)       &    0.0599  & --        & $<-10.5$     & --           & $<-10.8$       & -- \\ 


&  1101+384 (Mrk421)         &  0.0308    & $-10.0$     & $-10.6$     
& $-10.8$ & $<-10.5$ & $<-10.6$    \\
&           &      &      &      
& AIROBICC: &   $<-11.3$&     \\
& 1214+381 (MS 12143+38)     & 0.062      & $-12.7$     & $-11.2$     & --       &   $<-10.9$  & -- \\
& 1219+285 (ON 231)          & 0.102      & $-11.8$     & $-10.6$     & $<-10.9$ &   $<-10.9$  & -- \\
& 1404+286 (OQ 208)          &  0.0797    & $-12.8 $    & $<-10.7$    & --       & $<-10.8$      & -- \\
& 1514+004 (PKS 1514+00)     &  0.053     & $-11.9 $    & $<-10.8$    & --       &   $<-10.2$     & -- \\
& 1652+398 (Mrk501)          &  0.033     & $-10.3$     & $<-10.5 $   
& $-11.1$    & $<-10.9 $     & -- \\
&           &      &      &      
& AIROBICC: &   $<-10.8$&     \\
& 1727+502 (II~Zw~077)          & 0.055      & $-10.8$     & $<-10.7$    & $<-10.8$   &  $<-10.5 $   & -- \\
& 1807+698 (3C371.0)         & 0.0512     & $-12.0 $    & $<-10.5$    & --       & --          & -- \\
& 2200+420 (BLLac)           & 0.069      & $-10.9$     & $-10.3$   & $<-10.9$ & $<-10.6 $    & --  \\
& 2201+044 (PKS2201+04)      & 0.028      & $-11.9$     & $<-11.1$    & --       &  $<-10.3$    & -- \\
& 2209+236                   & --         &  --       & $-10.7 $    & --       &  $<-10.7$     & --  \\
\end{tabular} 
\end{minipage}
\end{table*} 
We compute the cascade spectra numerically as described in
Mannheim et al. (1991) and Mannheim (1993a), but taking
into account additionally 
Bethe-Heitler pair production and proton synchrotron
radiation assuming
a photo-pair luminosity $L_{\rm e^\pm}\simeq 0.5L_\pi$ 
(strictly valid
only for
straight $\alpha=1$ power law photon fields as a target;
Sikora et al. \cite{sikora87}).  
The optical depth of the jet with respect to
pair creation is computed as follows:
Since the typical angle of emission in the comoving frame
is $\theta'\approx \pi/2$, the intersection 
of the line of sight with the jet corresponds to
its transverse radius $r$ (denoted as
$r_\perp$ in Mannheim \cite{mannheim93}).  
The target radiation ($>$far-infrared)
becomes optically thin at 
$r=r_{\rm b}$ where we expect the largest $\gamma$-ray luminosity
to be produced (see caption of Fig.3).    
The comoving-frame optical depth is then given by
\begin{equation}
\tau_{\gamma\gamma}'=n_{\rm syn}'\sigma_{\gamma\gamma}r_{\rm b}=
a u_B \epsilon_\circ'^{-1}  \ln[\nu_{\rm c}/\nu_{\rm b}]^{-1}
\sigma_{\gamma\gamma}r_{\rm b}
\end{equation}
where
\begin{equation}
a={u_{\rm syn}\over u_{\rm B}}\simeq \gamma_{\rm j}
\beta_{\rm j}\phi\left(1+\eta{\Lambda_{\rm e}/ \Lambda_{\rm p}}\right)^{-1}
\end{equation}
(cf. Eq.(24) in Blandford \& K\"onigl \cite{blandford79}).
The pair-creation cross section
reaches $\sigma_{\gamma\gamma}\simeq {1\over 3}\sigma_{\rm T}$ at 
the resonant
target photon energy $\epsilon_\circ'=2(m_{\rm e}c^2)^2/E'$
for a $\gamma$-ray with energy $E'$.
Using the Lorentz invariance of the optical depth, i.e.
$\tau_{\gamma\gamma}(E)=\tau_{\gamma\gamma}'(E')$,
we obtain the optical depth in the observer's frame. 
Synthetic cascade spectra were produced with
the following input parameters:
redshift $z$, jet Lorentz factor $\gamma_{\rm j}$,
angle to the line of sight $\theta$, proton-to-electron
cooling rate ratio $\xi$, proton-to-electron energy density ratio $\eta$, jet
luminosity $L_{\rm jet}$ (magnetic field + relativistic particles),
synchrotron cutoff frequency in the comoving frame $\nu_{\rm c}'$.
While $\xi$ affects only the cascade part of the spectrum,
$\eta$ also affects the electronic
synchrotron spectrum for fixed $L_{\rm jet}$, since it  
imparts the
given energy between electrons and protons.
In general, the synchrotron and cascade spectrum are both rather
robust.  The most important effect is the dramatic variation of
the observed flux for variations with the boosting 
angle $\theta$.  Variations of the cosmological parameters
$\Omega$, $\Lambda$ and $H_\circ$ from their adopted
values 1, 0 and 75~km~s$^{-1}$~Mpc$^{-1}$ do not significantly affect
the model fits, they do affect
the luminosities in Tab.2 accordingly.
\section{Multifrequency modeling of nearby blazars}
\label{spectra}
\subsection{The sample} 
\label{sample} 
We investigated the spectra
of blazars from the list of Fichtel et al. (1994)
which have a redshift $z\le 0.1$ or unkown, and which
are visible from the northern hemisphere.
The sources typically have a flat or inverted radio spectrum $>1$~Jy.
We used published multifrequency data,  
ROSAT data from the all-sky survey as well as from public
pointings, preliminary flux limits from Whipple observations
and HEGRA scintillator and AIROBICC flux limits.
At the time of writing, 5 out of a total of 15 sources were
detected by EGRET, 2 = Mrk421 and Mrk501 were detected by Whipple 
(Punch et al. \cite{punch92},
Kerrick et al. \cite{kerrick95b}, Quinn et al. \cite{quinn96}).
Recently, the detection of Mrk421 has been confirmed independently
by the HEGRA $\check{\rm C}$erenkov telescopes (Petry et al. 1995).
Extending the sample to $z=0.2$ would
roughly double the number of sources.\\
To estimate the effect of intergalactic absorption,
we assumed $z=0.3$ for those sources, where no redshift
was available.  At $z<0.3$, the corresponding
host galaxies are generally expected to be  
seen.  We used the mean diffuse background density shown in Fig.2
for the computation of the absorption.
\\
The data sets are neither simultaneous nor complete,
and the intrinsic source 
variability, which shows up as a large dispersion of
the fluxes for a given photon energy (e.g. TeV outburst of
Mrk421 by a factor of $\sim10$; Kerrick et al. \cite{kerrick95b},
Macomb et al. \cite{macomb95}), brackets the
theoretical spectrum.   
Since dust-rich quasars such as 3C273 (Lichti et al. \cite{lichti95},
Mannheim \cite{mannheim93b})
are not present in our sample, additional absorption of
the $\gamma$-rays by infrared photons from the host
galaxy can be neglected.
Predictions of TeV fluxes are 
given within the above uncertainties which may well add up to
a factor of $\sim$a few.  Table 1 summarizes the high
energy measurements of the sources comprising our sample.
\subsection{Summary of results} 
\label{results} 
\begin{figure*}
\newcommand{\picturebox}{\framebox[18cm]{\rule{0pt}{22.5cm}}}
\iffigures
\centerline{\psfig{figure=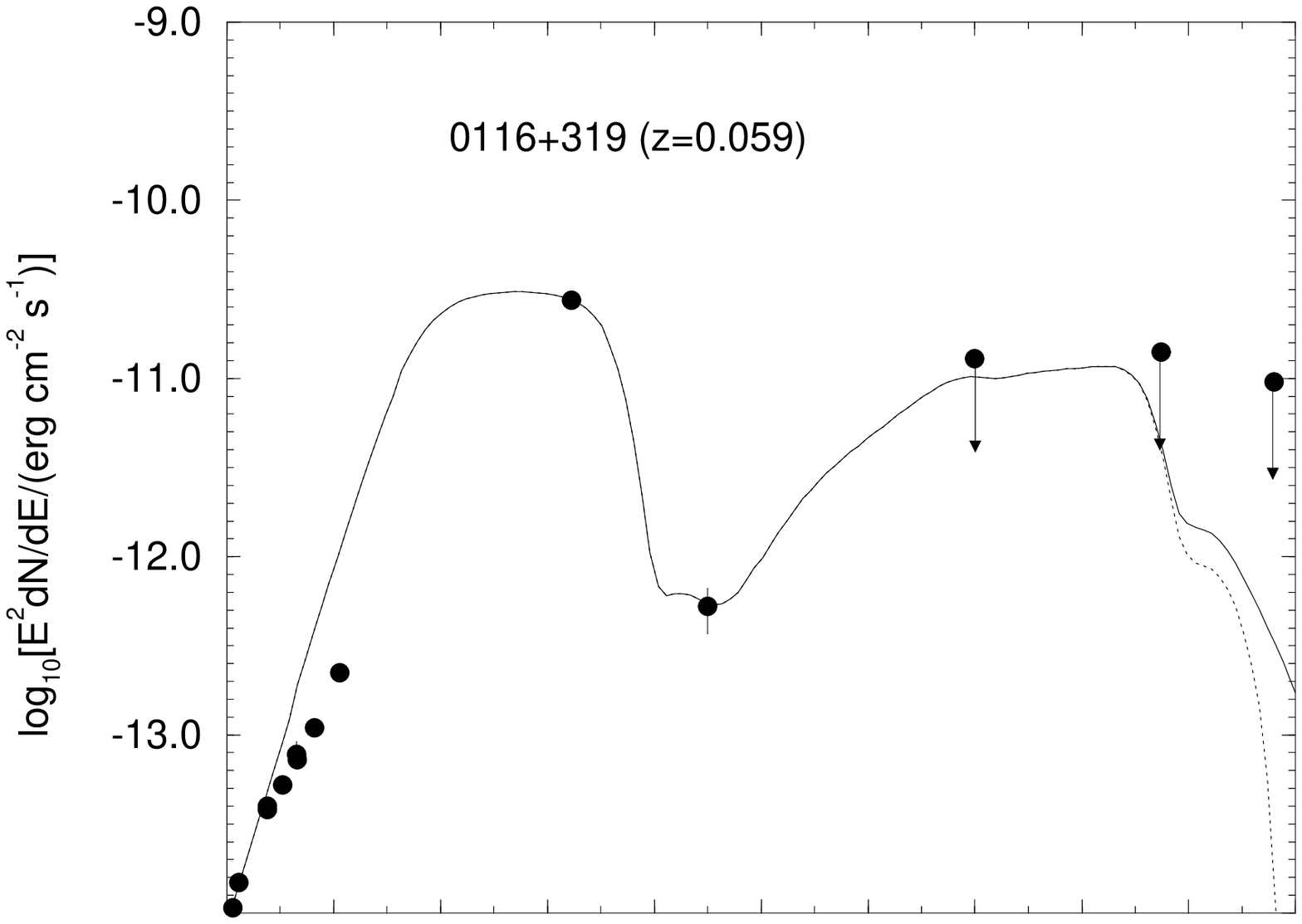,height=6.0cm}\hspace{-1.853cm}
            \psfig{figure=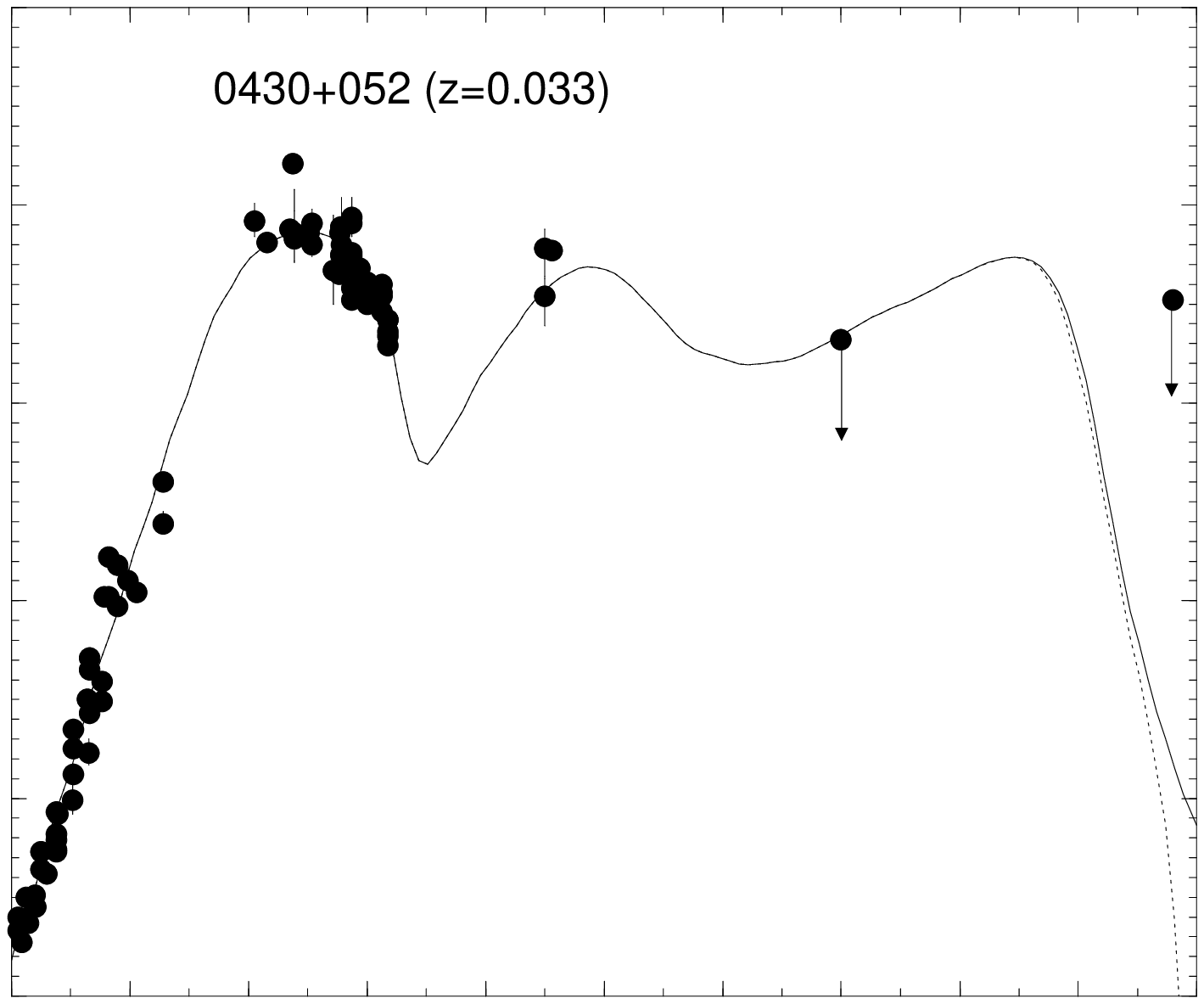,height=6.0cm}\hspace{-1.853cm}
	    \psfig{figure=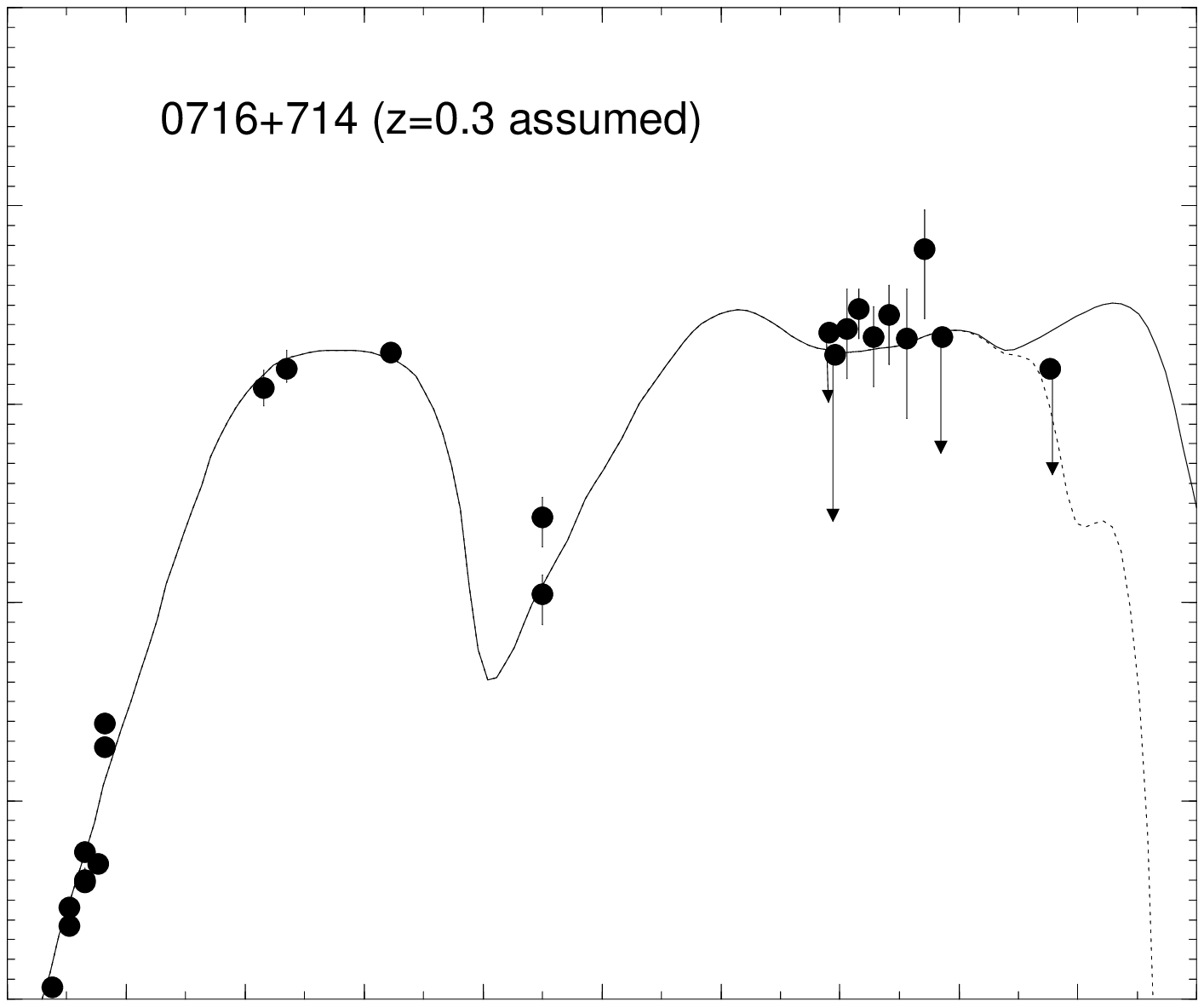,height=6.0cm}}\vspace{-1.48cm}
\centerline{\psfig{figure=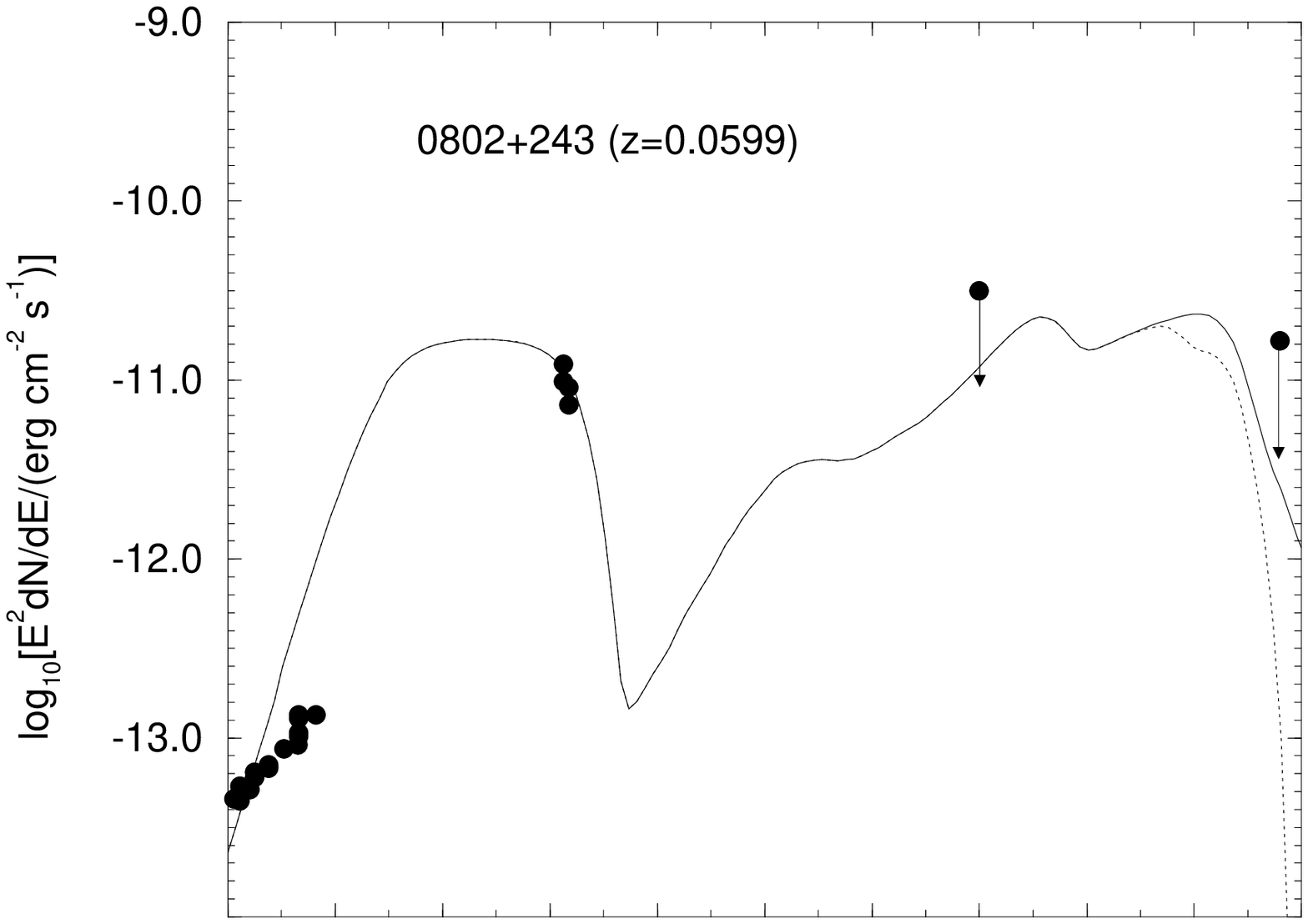,height=6.0cm}\hspace{-1.853cm}
            \psfig{figure=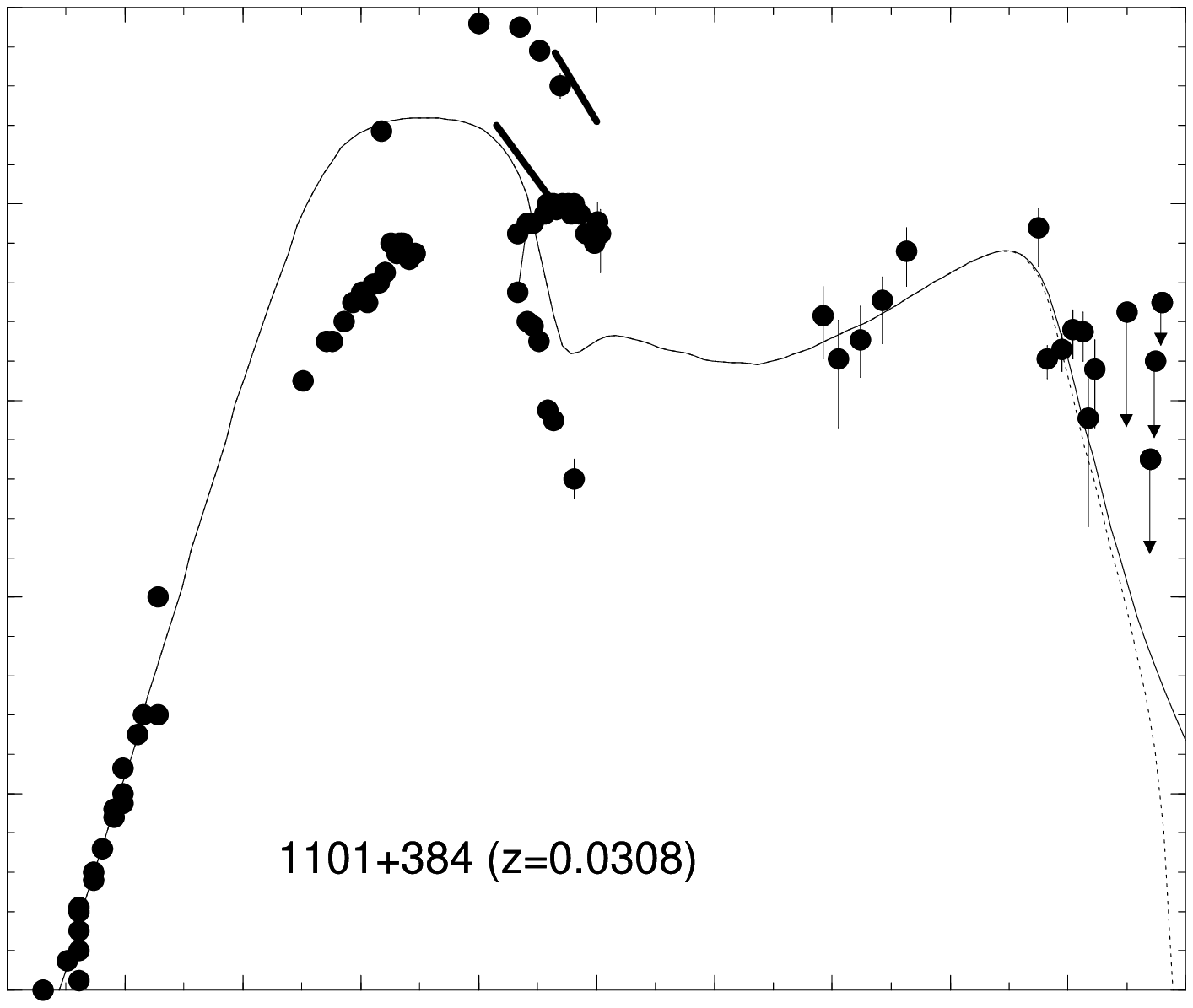,height=6.0cm}\hspace{-1.853cm}
            \psfig{figure=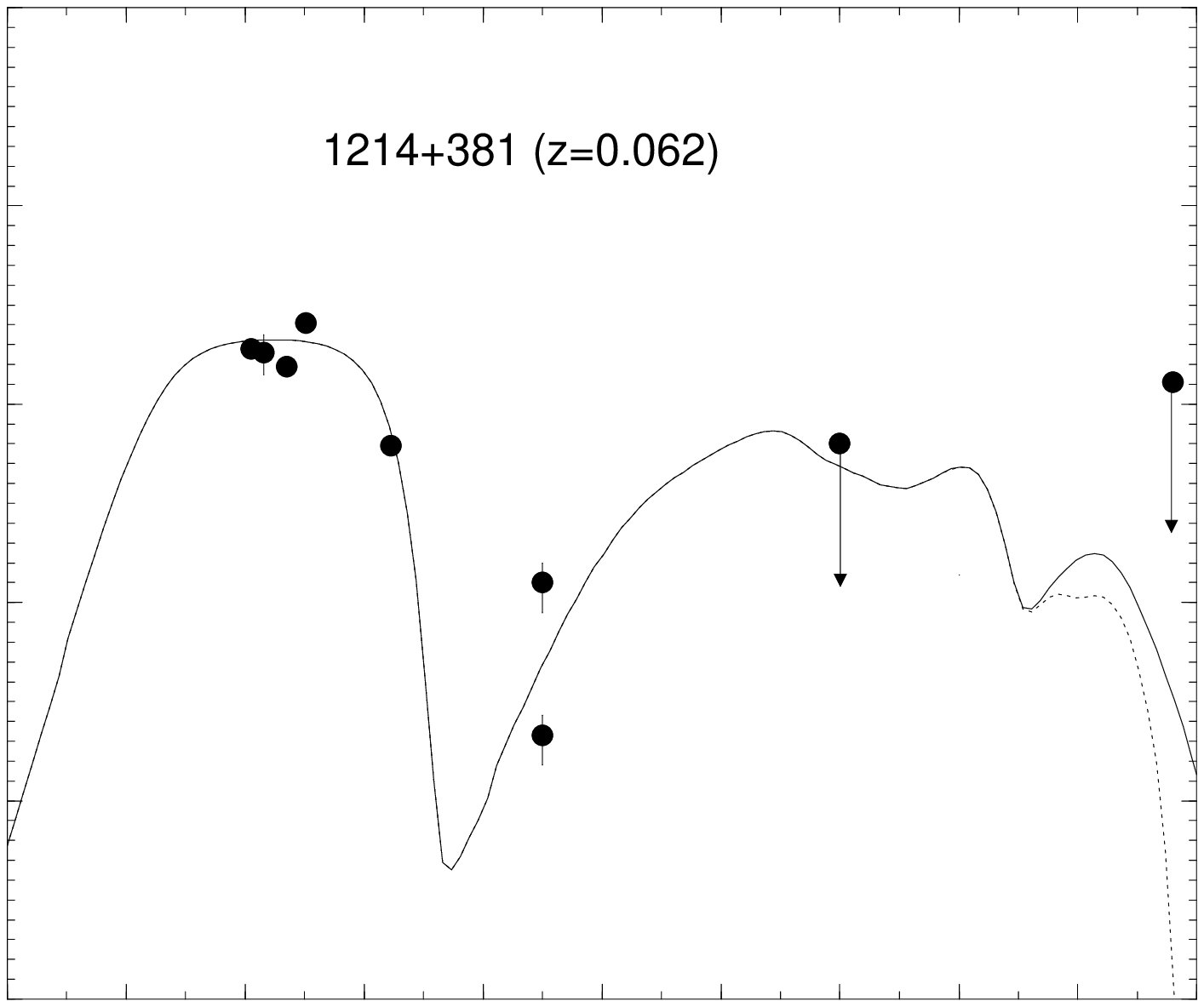,height=6.0cm}}\vspace{-1.48cm}
\centerline{\psfig{figure=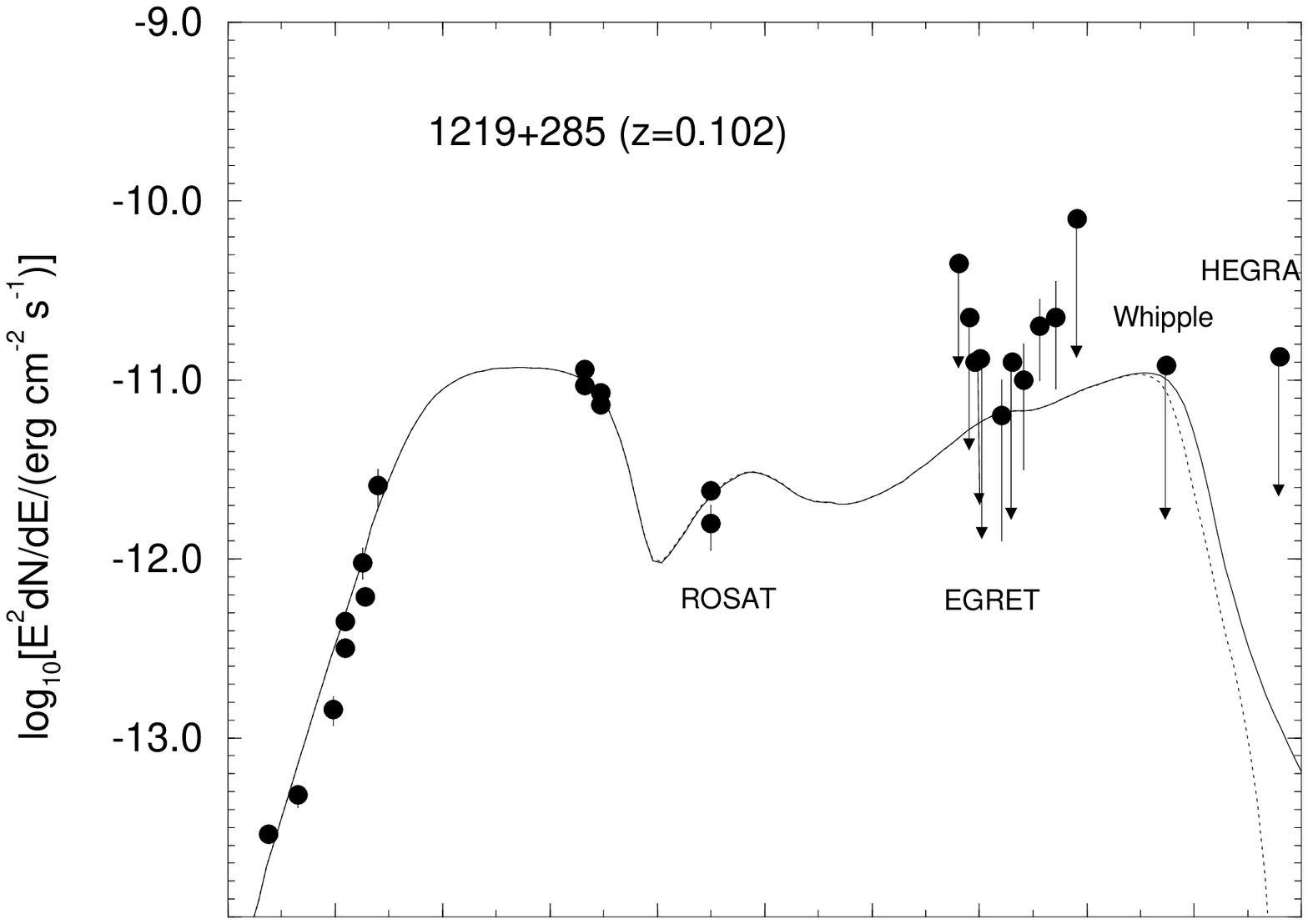,height=6.0cm}\hspace{-1.853cm}
            \psfig{figure=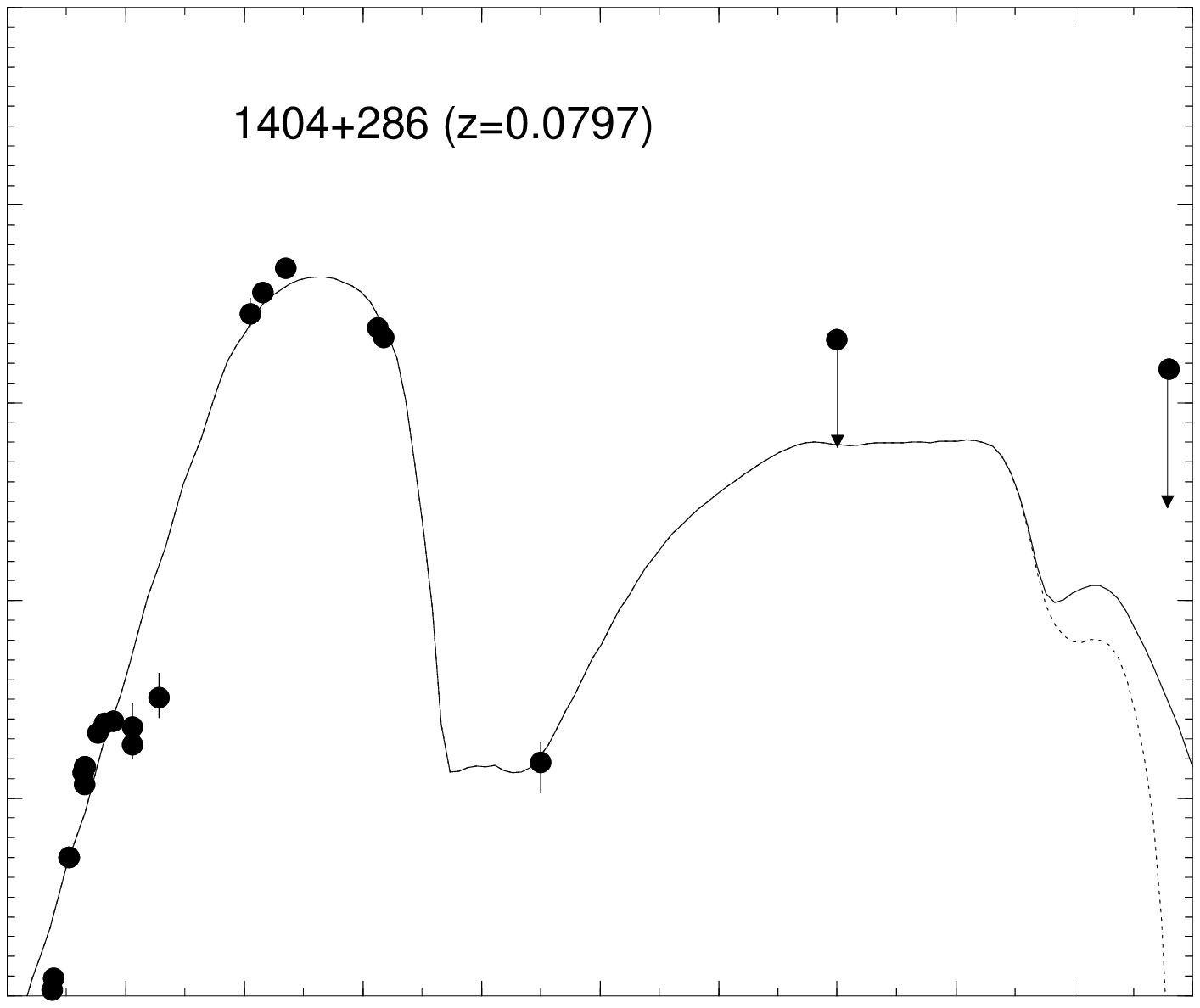,height=6.0cm}\hspace{-1.853cm} 
            \psfig{figure=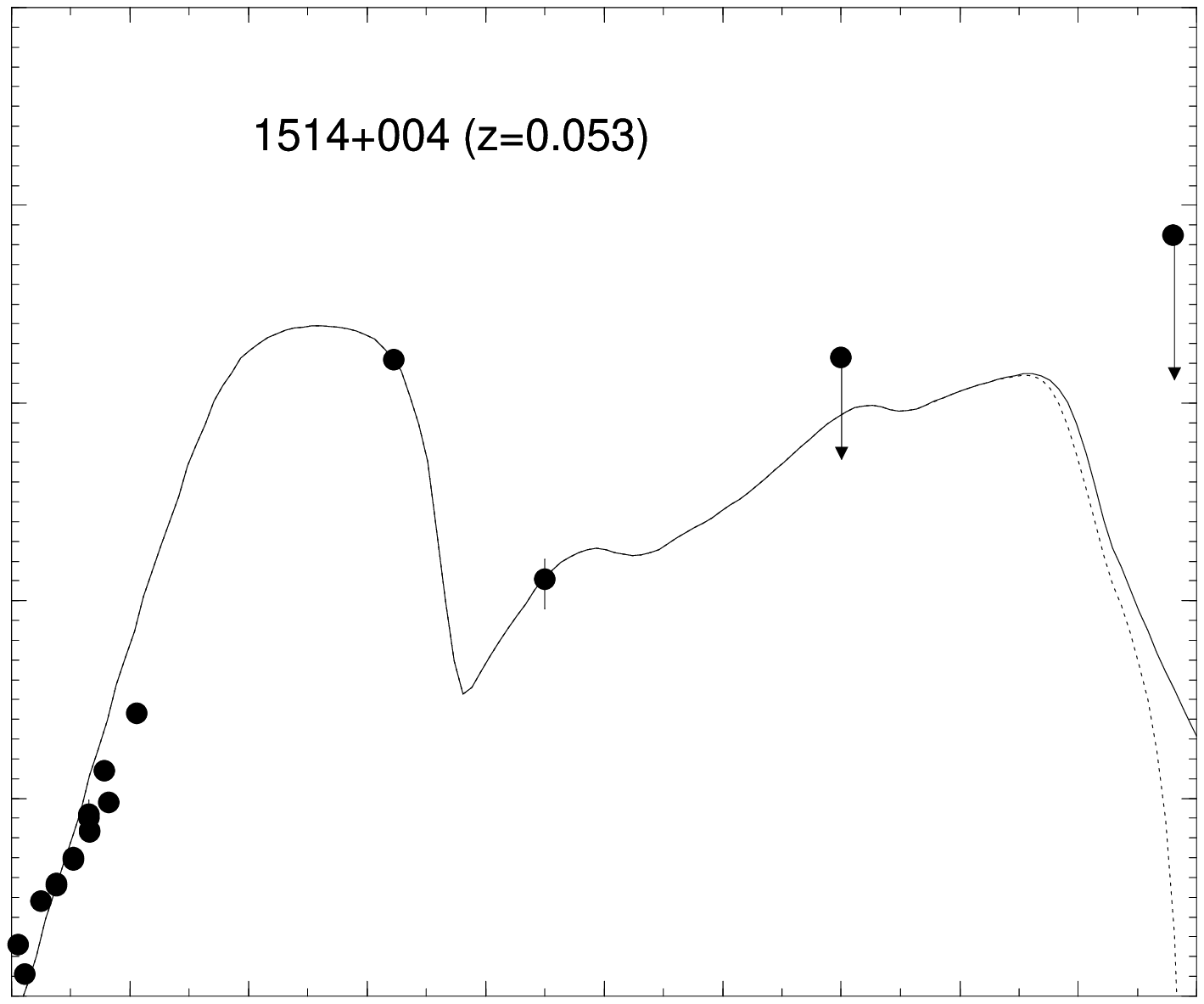,height=6.0cm}}\vspace{-1.48cm}
\centerline{\psfig{figure=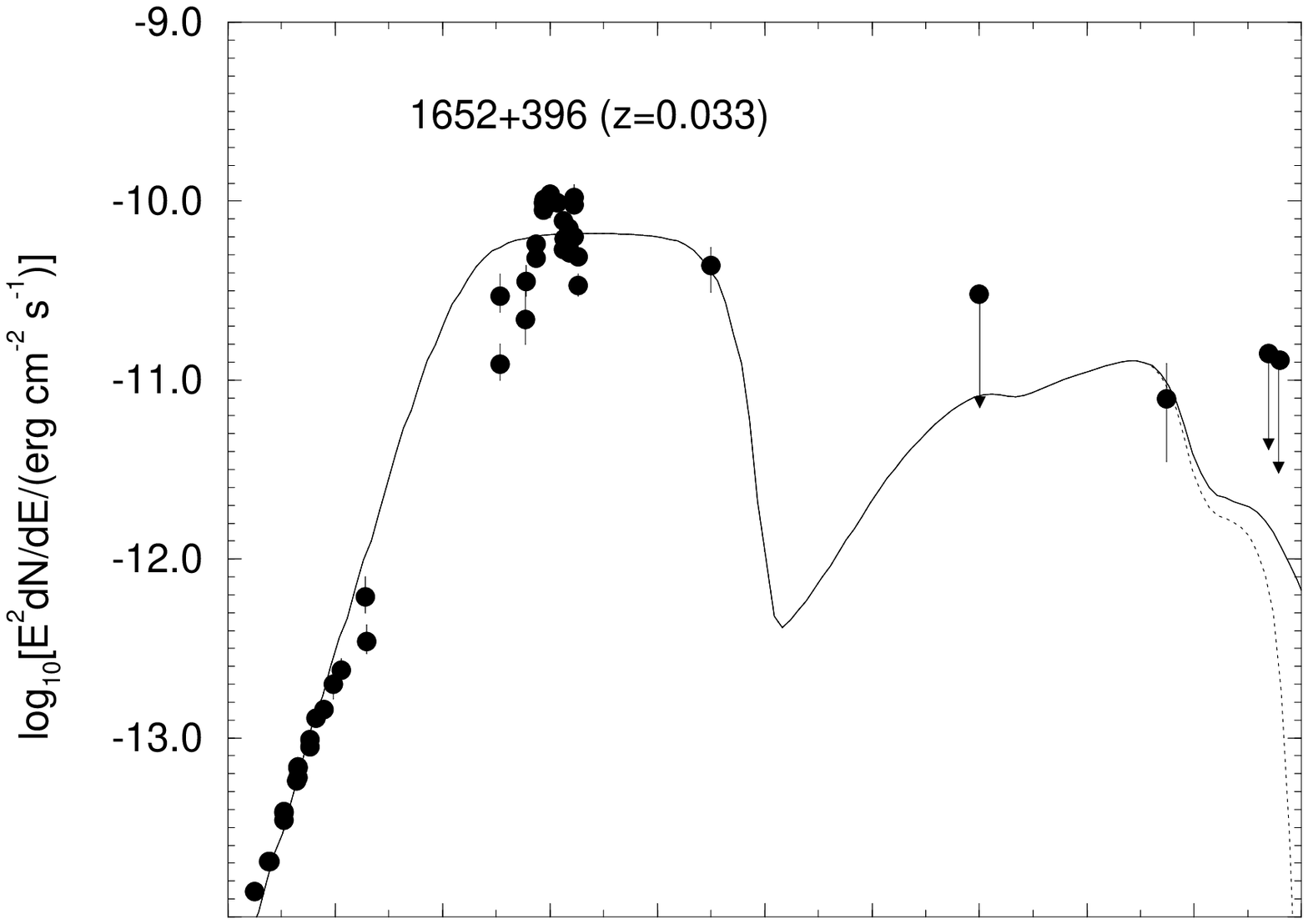,height=6.0cm}\hspace{-1.853cm} 
            \psfig{figure=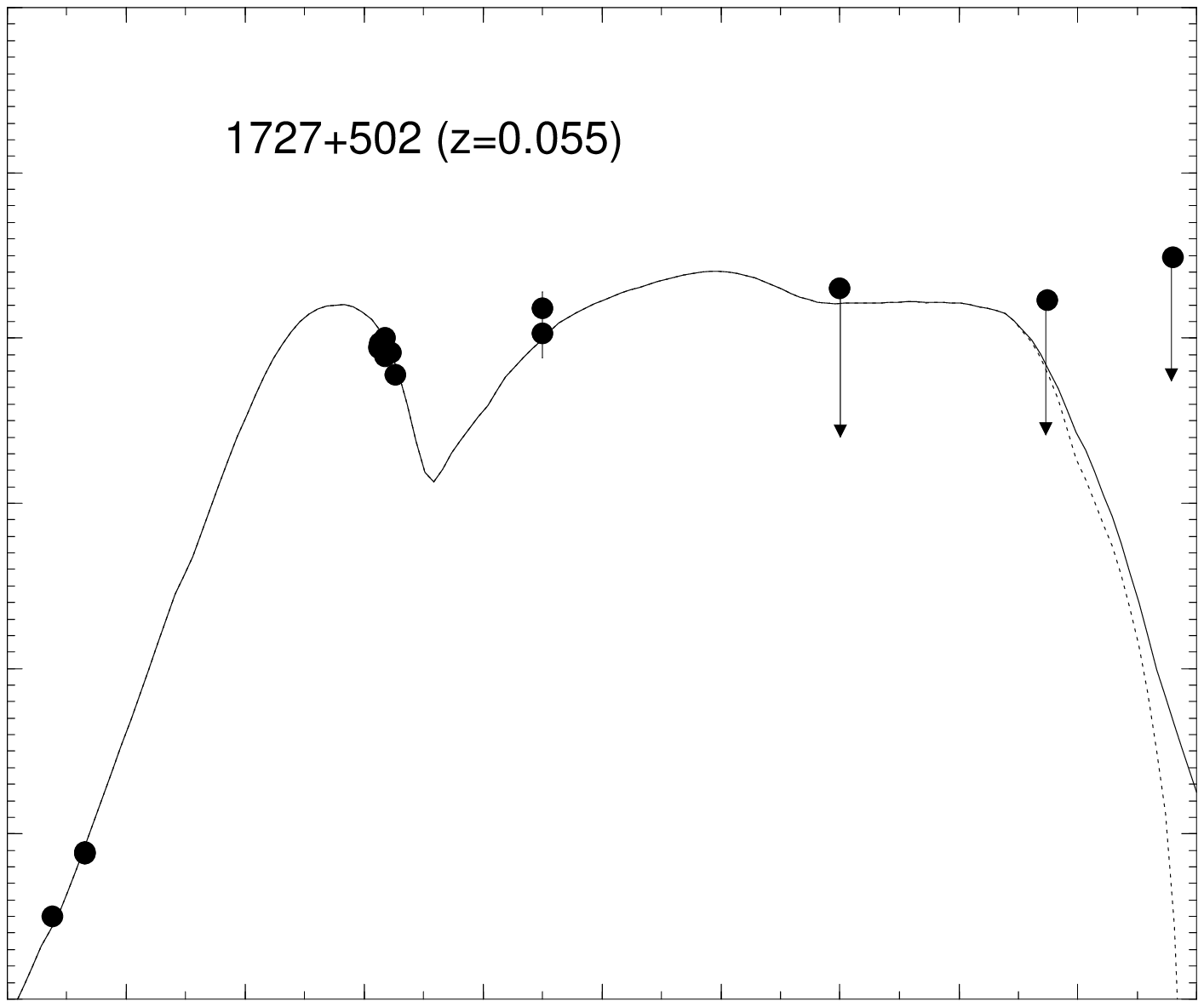,height=6.0cm}\hspace{-1.853cm} 
            \psfig{figure=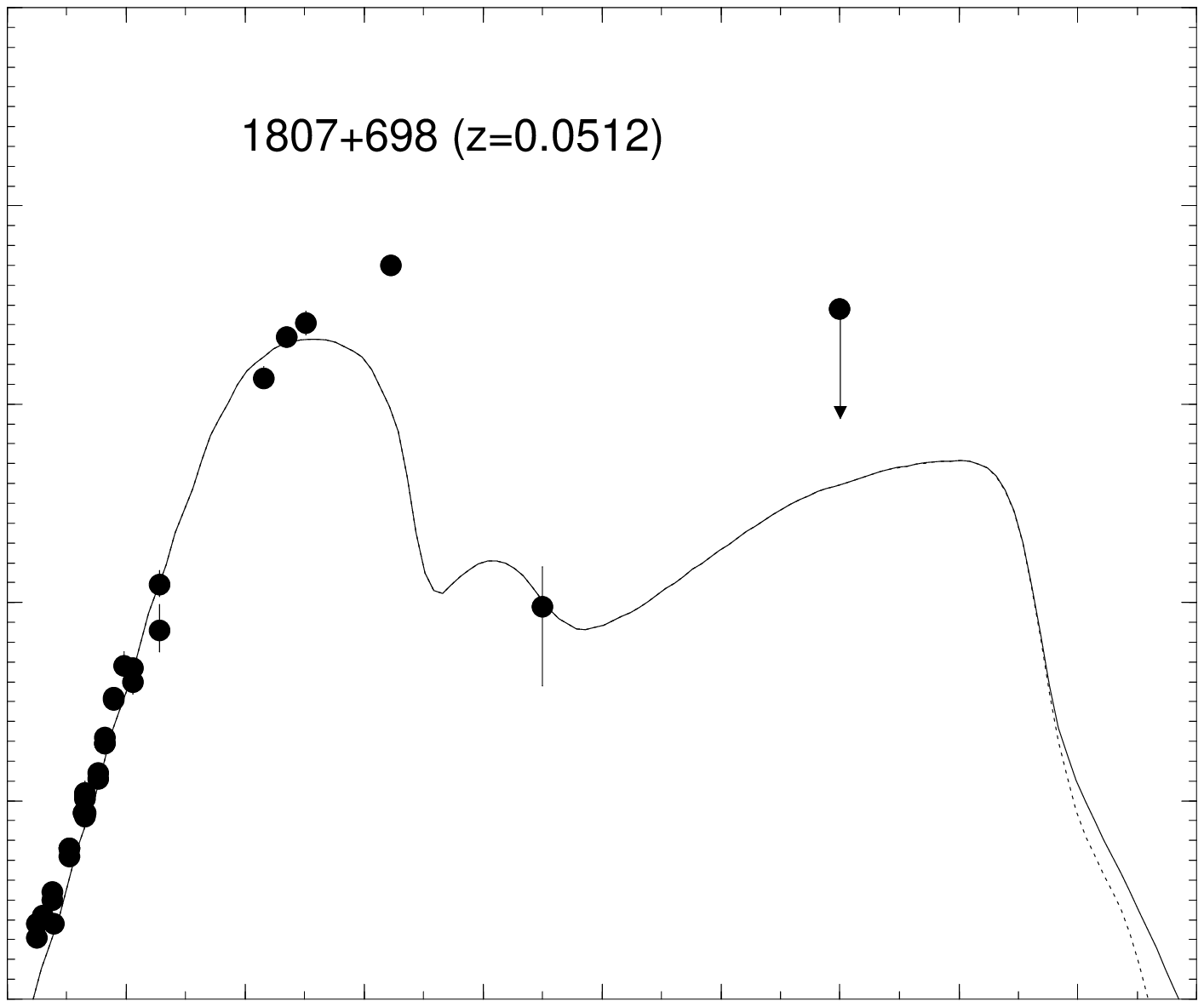,height=6.0cm}}\vspace{-1.48cm}
\centerline{\psfig{figure=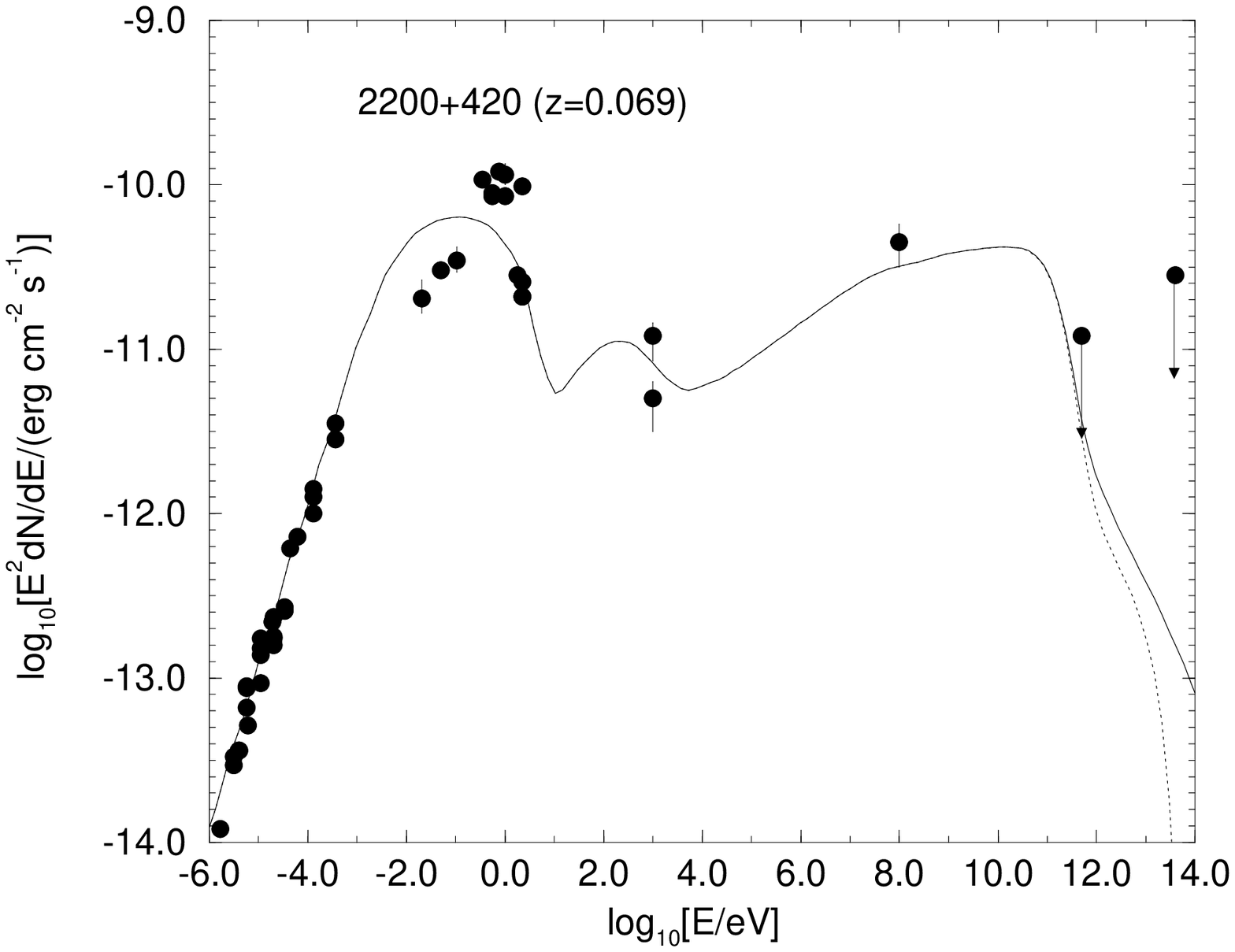,height=6.0cm}\hspace{-1.853cm} 
            \psfig{figure=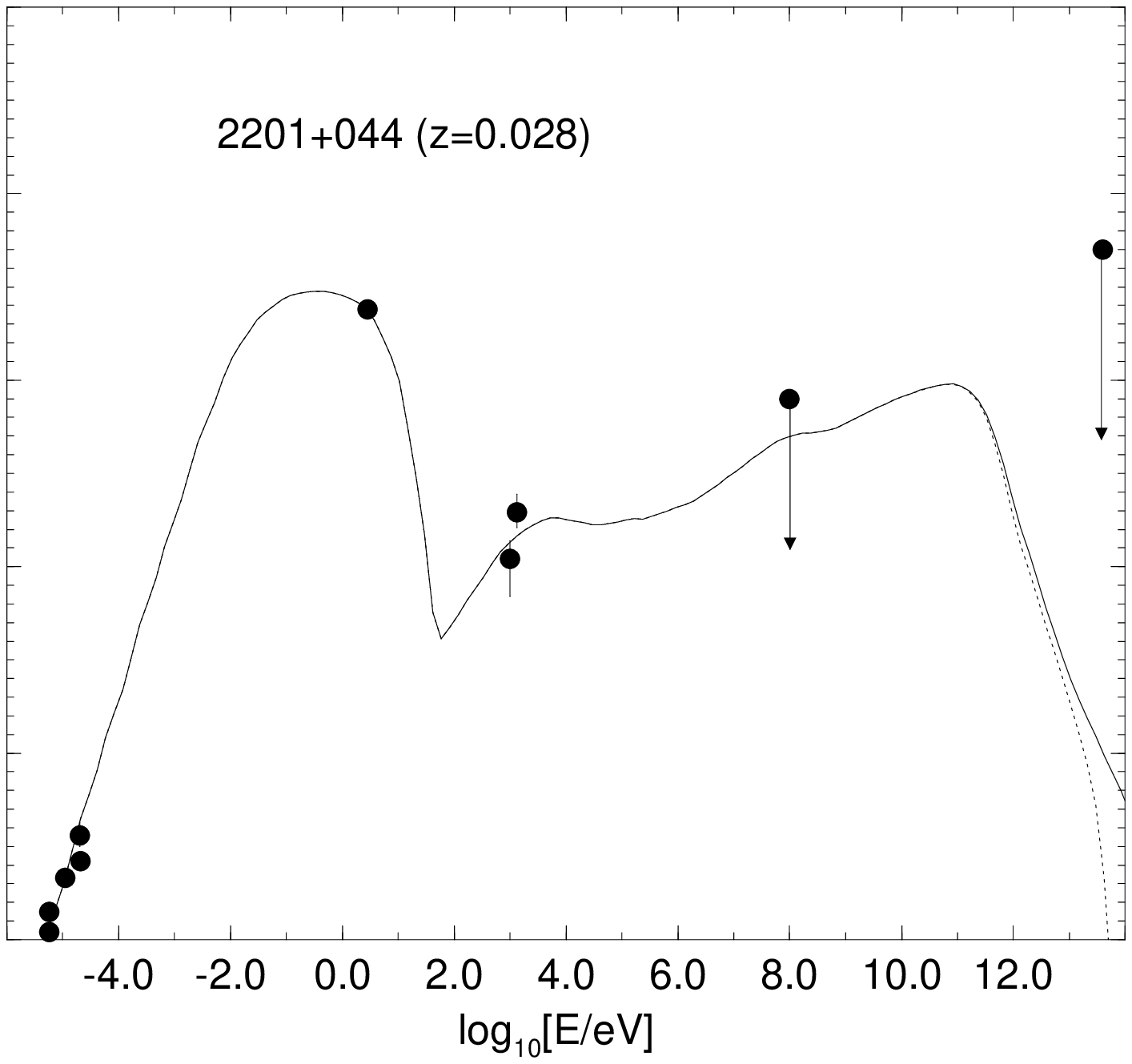,height=6.0cm}\hspace{-1.853cm} 
	    \psfig{figure=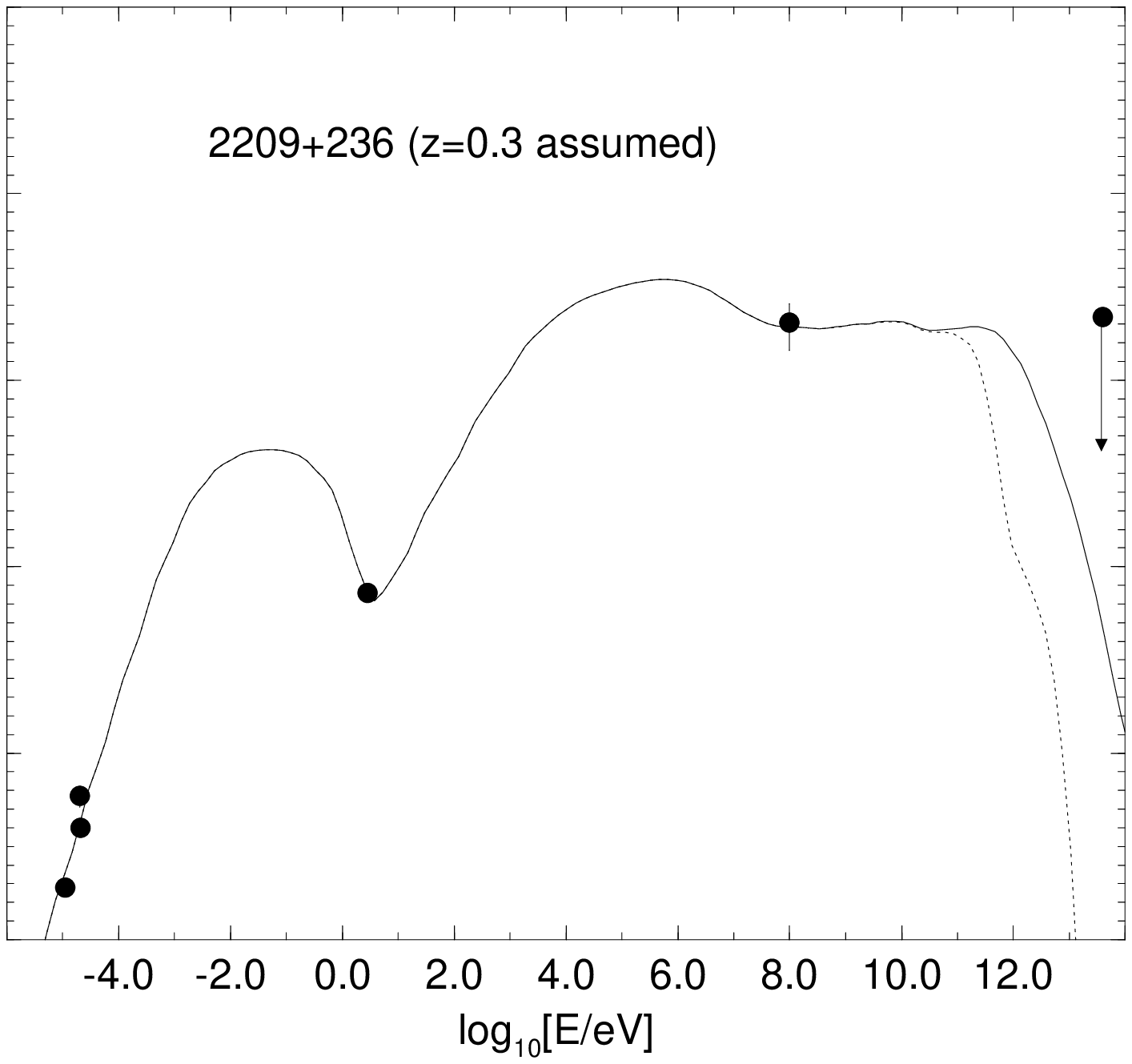,height=6.0cm}}            
\else     
\centerline{\picturebox}
\fi
\caption[]{{\it Solid lines:}
proton blazar model fits.  {\it Dotted
lines:} same model fits with external (cosmic) absorption taken into account}
\end{figure*}  

\begin{figure*}
\newcommand{\picturebox}{\framebox[14cm]{\rule{0pt}{11cm}}}
\iffigures
\centerline{           \psfig{figure=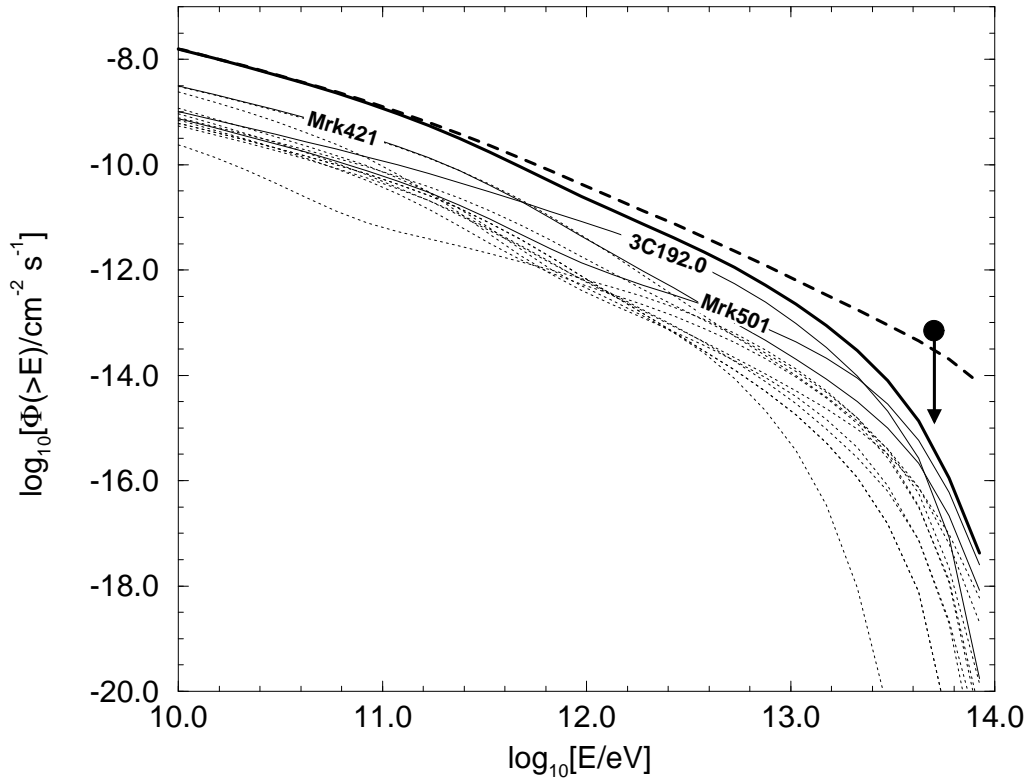,height=12cm}}
\else     
\centerline{\picturebox}
\fi
\caption[]{ 
Integral fluxes and cumulative HEGRA flux limit of the 13 blazars visible at La Palma  
(excluding 0716+714 and 1807+698) from 10~GeV to 100~TeV.  
{\it Dashed bold line:}  sum of fluxes with internal absorption.
{\it Solid bold line:} sum of fluxes with internal+external absorption}
\end{figure*}
The data used to obtain the model fits
are non-simultaneous and there are large gaps 
in frequency coverage.
Therefore, the parameter choices, 
and hence the spectra, are not free of ambiguities. 
The accuracy of the model 
predictions cannot 
be much better than to a factor of $\sim$a few.  
Nevertheless, some important results can be inferred from the model fits:\\
\noindent
{\it Gamma-to-optical luminosity ratios:}
Among the randomly sampled nearby blazars, there is no  
source with a strong $\gamma$-flare.
The $\gamma$-ray luminosities  
are 
$L_\gamma\approx \eta\xi L_{\rm opt}\approx L_{\rm opt}$
with
a proton-to-electron
ratio $\langle \eta\rangle=80\pm 30$
and $\langle\log_{10}[\xi]\rangle =-1.8\pm 0.5$.   
The absence of flares could be due to either a low duty cycle
for $\xi\approx 1$ states or, 
alternatively, 
the acceleration process could generally 
be less efficient in BL~Lacertae objects than in radio quasars  
(most nearby blazars in our sample are BL~Lacertae objects). 
This could be related to the magnetic
field structure which is preferentially perpendicular to the jet axis
in low-luminosity BL~Lacs and parallel in high-luminosity quasars (Bridle \&
Perley 1984).\\
\noindent
{\it TeV $\gamma$-rays:}  We note that promising candidates for detection
at TeV energies are 0802+243, 0430+052, 1514+004 and 1219+285
(in addition to Mrk421 and Mrk501).  The
TeV photon flux predicted for 0716+714 accounting for cosmic absorption
at $z=0.3$ is similar to that for 1219+285.  However,
we consider this redshift as a lower limit implying that the 
cosmic absorption is likely to be much stronger than assumed.\\
\noindent
{\it Spectral slope between TeV and 50~TeV:}
Above TeV,
the spectra continue as power laws, but with a steeper slope.
The intrinsic source spectrum 
is given by
$I_{\rm s}=I_\circ (1-\exp[-\tau_{\gamma\gamma,\rm int}])/
\tau_{\gamma\gamma,\rm int}$ 
corresponding to
absorption in a homogeneous source, for which the photospheric size 
is energy-dependent.  Hence, for $\tau_{\gamma\gamma,\rm int}\gg 1$
and since $\tau_{\gamma\gamma,\rm int}\propto E$,
the asymptotic spectrum is given by $I_{\rm s}\propto I_\circ /E$.
This is in marked contrast to the external absorption, which is
given by an exponential law, i.e. 
$I_{\rm obs}=I_{\rm s}\exp[-\tau_{\gamma\gamma, \rm ext}]$.\\
\noindent
{\it Similarity of comoving-frame spectra:}
Somewhat surprising, we find that the comoving-frame break
frequencies are roughly constant with $\nu_{\rm b}' 
\approx 10^{11}$~Hz.  The comoving-frame cutoff frequencies also agree
with a constant value of $\nu_{\rm c}'\approx 3\,   10^{14}$~Hz,
which is theoretically expected from non-relativistic
theory (Biermann \& Strittmatter \cite{biermann87}).  This could be a hint that
the physics in the comoving frame of the relativistic inner jet is 
basically the same as in the (non-relativistic) hot spots of the jet hundreds of
kiloparsecs away from the origin of the jet (Harris et al. 1994).\\
{\it  Proton maximum energy:}
The average proton maximum energy in the observer's frame inferred from the fits is
$\sim 10^{10}$~GeV.  Detection of the resulting neutrino background from 
blazars seems possible (Mannheim \cite{mannheim95}).\\
\noindent
{\it  Jet Lorentz factor:}
The average jet Lorentz factor $\langle\gamma_{\rm j}\rangle_{\rm obs}=10\pm 4$ is 
consistent with the value $\gamma_{\rm j}=7$
inferred for low-luminosity radio sources 
based on the unification of BL~Lacs with Fanaroff-Riley type I
galaxies (Urry \& Padovani \cite{urry95}).
A freely expanding jet should have $\Phi\approx \gamma_{\rm j}^{-1}\approx 6$~deg
which is consistent with the value from the fits.\\
\noindent
{\it Variability time scales:}
From the model fits and Eq.(4) we obtain the typical 
numbers  $\langle u_{\rm syn}/u_{\rm B}\rangle=0.01$, 
$\langle B\rangle =40$~G, $\langle r_{\rm b}\rangle =10^{15}$~cm
and the Doppler factor $\langle \delta \rangle =10$.
Using Eq.(3), we obtain the $\gamma$-ray break energy 
$\langle E\rangle =\langle\delta\rangle \langle E'\rangle=1$~TeV
where intrinsic absorption sets in ($\tau_{\gamma\gamma}'(E')=1$).
These numbers imply a minimum
variability time scale $\langle\Delta t_{\rm obs}\rangle
= \langle r_{\rm b}\rangle (1+z)/(\langle \delta\rangle c)
\approx 3\, 10^3$~s in the observer's frame.

\begin{table*}
\caption{Physical parameters for
the model fits shown in Fig.4 and  
predicted high-energy $\gamma$-ray fluxes.
The last four columns give 
the predicted integral $\gamma$-ray fluxes 
above $1$ and 40~TeV in units of $10^{-14}\rm cm^{-2}~s^{-1}$ 
accounting for  {\it internal} absorption 
(index i) and (ii) {\it internal + external} absorption (index a) 
adopting the same differential spectra as noted in Tab.1}
\begin{minipage}{\textwidth}
\renewcommand{\footnoterule}{\rule{0pt}{0pt}}
\renewcommand{\thefootnote}{\arabic{mpfootnote}}
\label{table}
\begin{tabular}{llccccccccccccc}
&\bf Source& $\gamma_{\rm j}$ & $\theta \over\rm deg$& 
$\phi\over \rm deg$& $\log_{10}\left[L_{\rm jet}\over{\rm erg/s}\right]$ &  
$\eta$& $\log_{10}[\xi]$ & $\log_{\rm 10}\left[\nu_{\rm c}' \over\rm Hz\right]$ & 
$F_{\rm i}(>{\rm TeV})$  & $F_{\rm a}(>{\rm TeV})$  
& $F_{\rm i}(>40\rm TeV)$  & $F_{\rm a}(>40\rm TeV)$\\
\hline
& 0116+319 & 6  & 6 & 7  & 44.8  &110 & -2.1 & 14.8 & 50 & 30  & 0.5 & $3 \,  10^{-2}$ \\
& 0430+052 & 8 & 8  & 5  & 45.0  & 30 & -1.3 & 13.8 & 600& 400 & 0.2 &  $5\,   10^{-2}$  \\
& 0716+714 & 16 & 2  & 1 & 44.5  & 70 & -1.3 & 14.4   & 900& 80  & 20  & $6\,   10^{-9}$ \\
& 0802+243 & 4  & 3  & 2 & 44.0  & 80 & -1.5 & 14.1  & 500& 500 & 4 & $2\,   10^{-1}$     \\
& 1101+384 & 20 & 2  &5.7& 45.2  & 50 & -2.2 & 15.8& 500& 400 & 0.6 & $1 \,   10^{-1}$    \\
& 1214+381 & 5  & 15 & 5 & 45.7  &100 & -2.2 & 14.2& 50 & 30 & 0.6 & $2\, 10^{-2}$     \\
& 1219+285 & 10 & 7  & 4 & 45.2  &100 & -1.8 & 14.8& 200& 80  & 0.2 & $7\,   10^{-4}$  \\
& 1404+286 & 10 & 3  & 4 & 44.7  &100 & -2.5 & 13.7&  40& 20  & 0.4 & $7\,   10^{-3}$  \\
& 1514+004 & 7  & 5 & 3.5& 44.5  &100 & -2.0 & 14.2 & 200& 200 & 0.6 & $4 \,   10^{-2}$    \\
& 1652+398 & 10 & 2  & 3 & 43.9  &100 & -2.5 & 16.5& 100 & 100  & 1.8 & $6 \,   10^{-2}$   \\
& 1727+502 & 10 & 5 & 10 & 44.0  & 30 & -1.1 & 13.8& 80 & 60  & 0.08& $6\,   10^{-3}$ \\
& 1807+698 & 10 & 9 &5.7 & 44.9  & 30 & -1.8 & 14.1& 4  & 3   & 0.02& $1\,   10^{-3}$  \\
& 2200+420 &  10 & 7& 10 & 45.6  &100 & -1.9 & 13.8& 50 & 30  & 0.2 & $6\,   10^{-3}$ \\
& 2201+044 &  10 & 5& 4  & 44.0  & 90 & -2.2 & 14.3 & 70 & 60  & 0.2 &$4 \,   10^{-2}$     \\
& 2209+236 &  10 & 7& 5  & 45.7  &100 & -0.8 & 13.6 & 400& 40  & 0.7 & $1\,   10^{-10}$ \\
\hline
& mean     
&  10     &   6   &   5    &   44.8   &   80     &   -1.8     &  14.4      & \\
& variance 
&   4     &   3   &   3    &   0.6   &   30     &   0.5     &   0.7     & \\ 
\end{tabular}
\end{minipage}
\end{table*}

\section{Current flux limits} 
In the TeV energy region, the search for point sources is only
possible using 
earth-bound experimental setups sensitive to the secondary particles
induced by cosmic ray primaries in the Earth's atmosphere.
The main task is to separate the air showers induced by primary
$\gamma$'s from the much more abundant hadronic showers. 
Apart from detecting the charged secondary particle component using a matrix
of scintillator counters, the photons of the \v{C}erenkov light cone 
produced by relativistic electrons also give valuable complementary
information about the shower and thus the incoming primary particle. 
Whereas up to now, scintillator arrays have provided only upper limits
for the flux from $\gamma$-ray point sources, the employment of imaging
\v{C}erenkov telescopes has been the most successful
ground based observation technique of the last years:
observation of ultra high energy
$\gamma$-radiation from the Crab nebula, Mrk421 and Mrk501 are reported. 
As an example for experimental setups searching for TeV $\gamma$-rays,
we quote the results obtained by the HEGRA detector array
located on the Canary Island La Palma (28.8 N, 17.7 W, 2200m a.s.l.).
As it comprises a scintillator array of 224 counters, 17 Geiger towers
for $\gamma$/hadron separation, and a $7\times  7$ matrix of open
\v{C}erenkov counters (AIROBICC),
HEGRA is able to equally exploit the information
provided by the \v{C}erenkov photons
and the charged particle component of air showers.
The observation of single objects is furthermore possible by the
use of 3 \v{C}erenkov telescopes
which form the first part of a system of 5 telescopes. 
The operation of AIROBICC is restricted to clear, moonless
nights, but the solid angle acceptance of about 1 sr allows to observe
a large number of sources simultaneously. 
As AIROBICC and scintillator arrays are sensitive to different
physical properties of air showers, they have different energy
thresholds: the scintillator array detects showers
with more than about 40 TeV energy of the primary particle,
the threshold for AIROBICC is at about 20 TeV.
As a consequence of the lower \v{C}erenkov light yield and
larger fluctuations in proton showers, the threshold energy for
proton induced showers exceeds the $\gamma$-shower thresholds by about
10 TeV.
The upper flux limits for $\gamma$-point sources from the AIROBICC array   
are of the order $10^{-13}\rm{cm^{-2}s^{-1}}$ with $\gamma$-shower energy
thresholds of about 25 TeV.
For the extragalactic sources Mrk421 and 
Mrk501 AIROBICC provides upper flux limits of 
$1.24\,   10^{-13}\,\rm{cm^{-2}\,s^{-1}}$
and $3.72\,  10^{-13}\,\rm{cm^{-2}\,s^{-1}}$
at threshold energies of $25.3$ and $24.0$ TeV respectively (Karle et al. 1995).
For the scintillator array, K\"uhn (1994) gives an upper flux limit
of $4.3\, 10^{-13}\,\rm{cm^{-2}\,s^{-1}}$ above 40~TeV for Mrk421.
The results quoted above were obtained without any attempt
to reduce the large
background of hadron induced showers, which are about $10^5$ times more
numerous than $\gamma$-induced showers.\\
Current data analysis applying cuts on the basis of $\gamma$/hadron-
separation techniques (Westerhoff et al. 1995a) yields a cumulative
flux limit of $7 \, 10^{-14}$~cm$^{-2}$~s$^{-1}$ (90\% CL)
above 50~TeV for the sum of the sources in our sample excluding
0716+714 and 1807+689 with poor visibility at La Palma (Westerhoff
\cite{westerhoff95b}). 
This is approximately equal to the sum of the {\it unabsorbed} 
theoretical fluxes  
derived from the model fits (Fig.5, Tab.2).  
Accounting for absorption, the theoretical flux drops by roughly
an order of magnitude lower.     It is difficult to assess the
systematical errors in the determination of the experimental
flux sensitivity, but they can be estimated to be quite large.
Since the theoretical errors are also difficult to control better
than within factors of $\sim $a few by the very nature of the
variable blazars, a detection of blazars at $\sim $50~TeV does not seem
beyond reach. 
On the other hand, very strong absorption, such
as proposed by Stecker et al. (\cite{stecker92})
and Dwek \& Slavin (\cite{dwek94}) indicated
as the dashed lines in Figs. 1 and 2, would further reduce the predicted
flux above 50~TeV by orders of magnitude.  In this case, detection
of blazars at HEGRA scintillator energies would be truly impossible. \\
Future analysis will include
methods to increase the signal-to-background-ratio by fully
exploiting the information of all components of the experimental setup
and by employing new and improved methods of data analysis.

\section{Conclusions}
\label{conclusions}
The continuum spectra of 15 nearby blazars 
have been modeled satisfactorily
over $\sim 20$ orders of magnitude in frequency
combining proton-initiated unsaturated synchrotron cascades and
electron synchrotron radiation emitted by relativistic
jets with a mean Lorentz factor $\langle\gamma_{\rm j}\rangle=10\pm 4$.  
The $\gamma$-ray spectra are complex,
but a mean power law photon index $s\approx 2$ represents
the spectra in the EGRET range rather well.
The inferred physical parameters are in line with a
proton-to-electron energy density ratio $\langle\eta\rangle=80\pm 30$.    
The proton cooling rate is less than
the electron cooling rate by a factor
$\sim 0.016_{-0.011}^{+0.034}$ implying about equal infrared-to-optical and 
$\gamma$-ray luminosities from the same emission volume.\\
Internal absorption of the $\gamma$-rays by
the low energy synchrotron photons leads to a steepening of the
spectra at $\sim $~TeV above which $s\approx 3$ up to $\sim 100$~TeV.  
The TeV spectra of Mrk421 and Mrk501 are not significantly affected
by cosmic absorption and should have
$s\simeq 3.2$.\\
Cosmic absorption {\it must} be present in 0716+714 
for which the predicted TeV emission without absorption greatly
exceeds the Whipple limit.
Cosmic
absorption becomes sufficiently strong for 0716+714
if its redshift is greater than 0.3.\\
Without
cosmic absorption, 
the sum of the blazar fluxes is about equal to
the current sensitivity of the HEGRA array.   Taking into account
cosmic absorption based on models of galaxy formation
by MacMinn \& Primack (1996), the predicted flux falls
an order of magnitude below the current flux limit.  
The absorption at $\sim 50$~TeV is not very sensitive to
cosmological parameters (e.g., $\Omega$ or $\Lambda$). 
A Hubble constant of $100$~km~s$^{-1}$~Mpc$^{-1}$ and  
late-epoch galaxy formation producing an infrared background
equal to that produced by IRAS galaxies would increase
our theoretical $\gamma$-ray fluxes by a factor of $\sim 3$.
A mild enhancement of the theoretical flux is 
expected from forward cascading of the pairs produced in
the intergalactic medium by inverse-Compton scattering off
background photons 
(Protheroe \& Stanev 1993, Aharonian et al. 1994, Plaga 1995), if the
intergalactic magnetic fields are weak.\\
A critical appraisal of the  systematic errors in 
the determination of the exact HEGRA energy threshold and flux level
as well as the  variable nature of the blazars 
leads us to conclude that  detection of the nearest
blazars with the HEGRA array seems feasible.   
A positive detection would disprove any claims of
an exceedingly strong cosmic
infrared background (with profound implications
on the era of galaxy formation) and would provide strong support
for a baryon-induced cascade origin of the $\gamma$-rays
from blazars.

\begin{acknowledgements} 
We are grateful to B. Hartman, the HEGRA collaboration,
C. von Montigny, J. Rose,
H.-C. Thomas and T. Weekes for their permission to use
preliminary results and for their comments.
We also thank our referee G. Setti for many helpful advices.
S.W. acknowledges support by the BMBF of Germany under grant 05 2WT 164.
This research has made use of the NASA/IPAC extragalactic database (NED).
\end{acknowledgements}
\noindent
{\footnotesize{\it Note added in proof:}
After submission of the manuscript, we became aware of a preprint
by Puget et al. (1996, A\&A, in press), 
claiming the detection of a diffuse FIR background using COBE data.
Within the uncertainties,
their measured flux is equal to the 
far-infrared flux of the mean MacMinn \& Primack background spectrum
used in our paper.}

\end{document}